\documentclass[a4paper,11pt]{article}
\pdfoutput=1

\usepackage{jheppub}

\usepackage{graphicx,array}
\usepackage{color}
\usepackage{latexsym}
\usepackage{amsthm}
\usepackage{amsmath}
\usepackage{amssymb}
\usepackage{hyperref}
\usepackage{bbold}
\usepackage{subfig}
\usepackage{multirow}
\usepackage{slashed}
\usepackage[normalem]{ulem}

\usepackage{hyperref}

\def\simgt{\mathrel{\lower2.5pt\vbox{\lineskip=0pt\baselineskip=0pt
           \hbox{$>$}\hbox{$\sim$}}}}
\def\simlt{\mathrel{\lower2.5pt\vbox{\lineskip=0pt\baselineskip=0pt
           \hbox{$<$}\hbox{$\sim$}}}}

\newcommand{\be}{\begin{equation}}
\newcommand{\ee}{\end{equation}}
\newcommand{\bea}{\begin{eqnarray}}
\newcommand{\eea}{\end{eqnarray}}

\newcommand{\GeV}{\textrm{ GeV}}
\newcommand{\TeV}{\textrm{ TeV}}
\newcommand{\SU}{\textrm{SU}}
\newcommand{\U}{\textrm{U}}
\newcommand{\Tr}{\textrm{Tr}}
\newcommand{\SM}{\textrm{SM}}
\newcommand{\gsim}{\lower.7ex\hbox{$\;\stackrel{\textstyle>}{\sim}\;$}}
\newcommand{\lsim}{\lower.7ex\hbox{$\;\stackrel{\textstyle<}{\sim}\;$}}

\newcommand{\cL}{\mathcal{L}}
\newcommand{\cM}{\mathcal{M}}

\newcommand{\GG}{\mathcal{G}}
\newcommand{\HH}{\mathcal{H}}

\newcommand{\ba} {\begin{eqnarray}}
\newcommand{\ea} {\end{eqnarray}}

\newcommand{\no} {\nonumber}
\newcommand{\cA} {\mathcal A}

\newcommand{\cLT} {{ \mathcal L}^{(T)}}
\newcommand{\cLS} {{ \mathcal L}^{(S)}}
\newcommand{\llq} {\lambda^q}
\newcommand{\lle} {\lambda^\ell}
\newcommand{\gq}{g_q}
\newcommand{\gl}{g_\ell}
\newcommand{\gls}{g_\ell^0}
\newcommand{\gqs}{g_q^0}

\font\tenrsfs=rsfs10 at 12pt

\font\sevenrsfs=rsfs7
\font\fiversfs=rsfs5
\newfam\rsfsfam
\textfont\rsfsfam=\tenrsfs
\scriptfont\rsfsfam=\sevenrsfs
\scriptscriptfont\rsfsfam=\fiversfs
\def\mathscr#1{{\fam\rsfsfam\relax#1}}

\font\tenbbm=bbm10 at 12pt
\newfam\bbmfam
\textfont\bbmfam=\tenbbm

\definecolor{darkblue}{cmyk}{1,0.3,0,0.2}
\definecolor{violet}{cmyk}{0,1,0,0.2}
\hypersetup{colorlinks, bookmarksnumbered, citecolor=darkblue, linkcolor=darkblue, pdfstartview=FitH, urlcolor=darkblue, linktocpage}

\newcommand{\arXhref}[1]{\href{http://arxiv.org/abs/#1}{#1}}

\preprint{ZU-TH-13/16}

\title{  Toward a coherent solution of diphoton and flavor anomalies}

\author[a]{Dario Buttazzo,}
\author[a,b]{Admir Greljo,}
\author[a]{Gino Isidori}
\author[a]{and David Marzocca}

\affiliation[a]{Physik-Institut, Universit\"at Z\"urich, CH-8057 Z\"urich, Switzerland}
\affiliation[b]{Faculty of Science, University of Sarajevo, Zmaja od Bosne 33-35, \\ 71000 Sarajevo, Bosnia and Herzegovina}

\abstract{We propose a coherent explanation for  the 750~GeV diphoton anomaly
and the hints of deviations from Lepton Flavor Universality in $B$ decays 
in terms a new strongly interacting sector with vectorlike confinement. 
The diphoton excess arises from the decay of one of the pseudo-Nambu-Goldstone bosons 
of the new sector, while the flavor anomalies are a manifestation of the exchange of the 
corresponding vector resonances (with masses in the 1.5--2.5~TeV range).
We provide explicit examples (with detailed particle content and group structure) of the 
new sector, discussing both the low-energy flavor-physics phenomenology and the signatures at high $p_T$. 
We show that specific models can provide an excellent fit to all available data.
A key feature of all realizations is a sizable broad excess in 
the tails of $\tau^+ \tau^-$ invariant mass distribution in $p~p \to \tau^+\tau^-$, 
that should be accessible at the LHC in the near future.}

\begin{document}

\maketitle
\flushbottom

\section{Introduction}
\label{sec:intro}

We propose a possible explanation of the diphoton anomaly~\cite{ATLAS:2016diphoton,CMS:2016owr} (see also Refs.~\cite{CMStalk,ATLAStalk,CMS:2014onr,Aad:2015mna})
and the hints of deviations from Lepton Flavor Universality (LFU) observed 
in $B\to D^{(*)} \ell \nu$~\cite{Lees:2013uzd, Huschle:2015rga,Aaij:2015yra}, $B \to K \ell^+ \ell^-$~\cite{Aaij:2014ora}, and $B \to K^* \ell^+ \ell^-$~\cite{LHCb:2015dla} decays in terms of a single coherent picture.
The basic idea is quite simple: we show that these two anomalies could be two correlated 
evidences of a new strongly interacting sector that confines and breaks spontaneously 
a large global symmetry around the TeV scale. 
Similarly to QCD, the lowest lying states of the spectrum are pseudo-Nambu-Goldstone bosons (pNGB)
and vector mesons. The diphoton excess arises from the decay of one of these pNGB, via its anomalous coupling to photons and gluons. The flavor anomalies are a manifestation of the exchange of the vector mesons,
under the assumption that these new states have a preferred coupling to SM  quarks and leptons of the  
third generation. 

The idea of a new strongly-coupled sector with vectorlike fundamental fermions charged also under the SM gauge group 
is also known in the literature as \emph{vectorlike confinement}~\cite{Kilic:2009mi,Kilic:2010et}.
The interpretation of the $750 \GeV$ diphoton excess as due to a pNGB in this context has already been widely 
discussed in the literature~\cite{Franceschini:2015kwy,Harigaya:2015ezk,Nakai:2015ptz,Bian:2015kjt,Craig:2015lra,Bai:2015nbs,Harigaya:2016pnu,Redi:2016kip,Kamenik:2016izk,Harigaya:2016eol}, as well as in Composite Higgs \cite{Buttazzo:2015txu,Low:2015qep,Bellazzini:2015nxw,Belyaev:2015hgo,No:2015bsn} or Technicolor \cite{Molinaro:2015cwg,Matsuzaki:2015che,Molinaro:2016oix} setups.
Particularly relevant to our setup is the extended discussion of such models presented in Ref.~\cite{Redi:2016kip}.
As we show, with a specific (and perfectly natural) 
choice of the techni-quarks, the vector resonances expected in this framework are 
ideal candidates for the vector field mediators postulated in Ref.~\cite{Greljo:2015mma} 
(a color-singlet $\SU(2)_L$-triplet vector)  and in Ref.~\cite{Barbieri:2015yvd} 
(a vector leptoquark) in order to explain the reported evidences of LFU violation in $B$ decays.
As in Refs.~\cite{Greljo:2015mma,Barbieri:2015yvd}, the observed flavor pattern of the 
$B$-physics anomalies is well reproduced under the assumption that the coupling of the 
composite vectors to SM fermions is controlled by an approximate $\U(2)^5$ 
flavor symmetry~\cite{Barbieri:2011ci}, that allows only couplings to SM fermions of the third generation in the limit of unbroken symmetry.

It is worth to stress that in our construction the mediators responsible for the 
flavor anomalies are unavoidable manifestations of the same dynamics responsible for the 
750~GeV diphoton excess (a similar goal, albeit limited to the neutral-current flavor anomalies, 
is pursued in Refs.~\cite{Belanger:2016ywb,Goertz:2015nkp}).
In this respect, the link between diphoton and flavor-physics anomalies
presented in our set-up is qualitatively different than those discussed 
in Refs.~\cite{Bauer:2015boy,Murphy:2015kag,Deppisch:2016qqd}.

In the minimal setups presented in this work the Higgs is an elementary scalar and the hierarchy problem of its mass is not solved. This point, however, can be addressed by enlarging the field content of the new sector, such that also the Higgs doublet arises as a pNGB, with the necessary custodial symmetry protection embedded in the unbroken chiral symmetry, as in usual composite Higgs models. Since the main connection between the diphoton excess and the flavor anomalies would not be qualitatively modified, in this work we focus on the minimal scenarios able to explain these experimental anomalies, postponing the study of a composite Higgs setup to future work.

The paper is organized as follows: in Section~\ref{sec:setup} we discuss the basic set-up,
with particular attention to the general aspects of the pNGB dynamics responsible for the 
diphoton excess. In Section~\ref{sec:models}
 we present two explicit models that fulfill the necessary ingredients
to explain both sets of anomalies.  In Section~\ref{sec:Flavor-mixing} we analyze the 
mixing between composite states and SM fermions, with particular attention to the 
flavor structure. A detailed discussion of the low-energy flavor-physics phenomenology, 
with a global fit of the free parameters in the explicit model with minimal particle content is
presented in Section~\ref{sec:flavor}. The main features of the high-$p_T$ phenomenology 
of the new states is discussed in Section~\ref{sec:LHCpheno}. The results 
are summarized in the Conclusions.

\section{General set-up}
\label{sec:setup}

As in Refs.~\cite{Franceschini:2015kwy,Harigaya:2015ezk,Nakai:2015ptz,Bian:2015kjt,Craig:2015lra,Bai:2015nbs,Harigaya:2016pnu,Redi:2016kip,Kamenik:2016izk,Harigaya:2016eol}, we add a new non-abelian gauge interaction with symmetry group $\SU(N_{TC})$ which confines at a scale $\Lambda = O(1 \TeV)$, and a set of elementary vectorlike fermions (denoted techni-quarks or TC quarks in the following) in the fundamental representation of this group and in some representations of the SM gauge group $\GG_{\SM}$. The number and type of SM representations define the number  $N_F$ 
 of independent techni-quarks, from the point of view of the new strong dynamics, and 
 the corresponding approximate non-anomalous global symmetry  $\GG = \SU(N_F)_L \times \SU(N_F)_R \times \U(1)_V$. This symmetry is explicitly broken by the masses of the fundamental techni-quarks, which we assign independently to each SM irreducible representation, and by the SM gauging itself.
We assume this theory condenses, like QCD, such that $\langle \bar{\psi}_i \psi_j \rangle = - f^2 B_0 \delta_{ij}$, breaking the (techni) flavor symmetry to the diagonal subgroup $\HH = \SU(N_F)_V \times \U(1)_V$. The SM gauge group is contained in this vectorial global symmetry, so that EW symmetry is preserved by the strong condensate.
This symmetry-breaking pattern provides $N_F^2 - 1$ pseudo-Nambu-Goldstone bosons (pNGB), classified in some $\GG_{\SM}$ representation.

\subsection{General aspects of pNGB dynamics} 

The low-energy dynamics of the pNGB is described by an effective chiral Lagrangian
completely analogous to the QCD one. The leading two-derivatives operator and the mass terms are\,\footnote{In the QCD case the value of $B_0$ can be extracted from the ratio between the pion mass, $m^2_\pi = B_0 (m_u+m_d)$, and the quark masses, obtaining $B_0 \approx 20 \times f_\pi$.}
\be
	\cL^{\chi_{PT}} = \frac{f^2}{4} \left\{ \Tr\left[ (D_\mu U)^\dag (D^\mu U)\right] + 2 B_0 (\Tr[\cM U] + \Tr[\cM^\dagger U^\dagger] ) \right\}~,
	\label{eq:1}
\ee
where $f$ is the pNGB decay constant, $\cM$ is the fundamental fermion matrix in flavor space, and $U(x) = \exp \left( 2 i \frac{\pi^a(x)}{f} t^a \right)$ is the pNGB matrix, transforming under $\GG$ as $U \to g_R U g_L^\dagger$. The $\SU(N_F)$ generators $t^a$ are normalized such that $\Tr[t^a t^b] = \frac{1}{2} \delta^{ab}$.

The neutral pNGB can have a coupling to two photons and gluons via the anomaly. In terms of the chiral field $U$ these are all included in the Wess-Zumino-Witten term, however for our purposes it is enough to write the relevant couplings as (see also \cite{Redi:2016kip})
\be
	\cL^{\rm WZW} \supset - \frac{g_b g_c }{16 \pi^2} \frac{\pi^a}{f} A_{V^b V^c}^{\pi^a} \, F^{b}_{\mu\nu} \widetilde{F}^{c\mu\nu}~,
\ee
where the anomalous coupling is given by $A_{V^b V^c}^{\pi^a} = 2 N_{TC} \Tr\left[ t^a t^b t^c \right]$, $t^a$ is the $\SU(N_F)_V$ generator corresponding to the $\pi^a$, $V^{b,c}$ and $g_{b,c}$ represent the SM gauge fields and couplings embedded in the same group, and finally $\widetilde F^b_{\mu\nu} = \frac{1}{2} \epsilon_{\mu\nu\rho\sigma} F^{b\rho\sigma}$.
We take one of the neutral pions coupled to both gluons and photons, in particular a singlet $\eta$ of $\GG_{\SM}$,\footnote{Neutral pNGB in non-singlet representation of $\SU(2)_L$ have necessarily a vanishing anomalous coupling to gluons, due to the traceless structure  of the $\SU(2)_L$ generators.} to fit the $750 \GeV$ diphoton excess.
Specifying the anomalous couplings to this case one has
\be
	\cL^{\rm WZW} \supset - \frac{\eta}{16 \pi^2 f} \left( g^{\prime 2} A_{BB}^{\eta} B_{\mu\nu} \widetilde{B}^{\mu\nu}
				+ g^{2} A_{WW}^{\eta} W^i_{\mu\nu} \widetilde{W}^{i\mu\nu}
				+ g_s^2 A_{gg}^{\eta} G^A_{\mu\nu} \widetilde{G}^{A\mu\nu} \right)~,
\ee
where
\be
	A_{BB}^{\eta} = 2 N_{TC} \Tr\left[ t_\eta Y^2 \right]~, \quad
	A_{WW}^{\eta} \delta^{ij} = 2 N_{TC} \Tr\left[ t_\eta \tau^i \tau^j \right]~, \quad
	A_{gg}^{\eta} \delta^{AB} = 2 N_{TC} \Tr\left[ t_\eta T^A T^B \right]~, \quad
\ee
and $Y$ is the diagonal $N_F \times N_F$ matrix with the hypercharges of the fundamental fermions, while $T^A$ and $\tau^i$ are the generators of $\SU(3)_c$ and $\SU(2)_L$ embedded in $\SU(N_F)_V$.
In terms of couplings with the gauge bosons mass eigenstates, these translate to
\be
	\cL^{WZW} \supset - \frac{e^2}{16 \pi^2} \frac{\eta}{f} \left(
			 A_{\gamma\gamma}^{\eta} F_{\mu\nu} \widetilde{F}^{\mu\nu}
			+ 2 A_{Z\gamma}^{\eta} F_{\mu\nu} \widetilde Z^{\mu\nu}
			+ A_{ZZ}^{\eta} Z_{\mu\nu} \widetilde Z^{\mu\nu}
			+ \frac{2}{\sin^2 \theta_W} A_{WW}^{\eta} W^+_{\mu\nu} \widetilde W^{-\mu\nu}
			\right),
\ee
with
\be\begin{split}
	&A_{\gamma\gamma}^{\eta} = A_{BB}^{\eta} + A_{WW}^{\eta} = 2 N_{TC} \Tr\left[ t_\eta Q^2 \right]~, \\
	&A_{Z\gamma}^{\eta} = A_{WW}^{\eta} \cot \theta_W - A_{BB}^{\eta} \tan \theta_W , \quad
	A_{ZZ}^{\eta} = A_{WW}^{\eta}  \cot^2 \theta_W + A_{BB}^{\eta} \tan^2 \theta_W ~.
\end{split}\ee
The decay widths to pairs of photons and gluons are then readily obtained as
\be
	\Gamma_{\eta \to \gamma\gamma} = \frac{\alpha^2}{64 \pi^3} |A_{\gamma\gamma}^{\eta}|^2\frac{m_{\eta}^3}{f^2}~, \qquad
	\Gamma_{\eta \to gg} = \frac{\alpha_s^2}{8 \pi^3} |A_{gg}^{\eta}|^2 \frac{m_{\eta}^3}{f^2}~,
\ee
and the ratios of other diboson decay widths to the diphoton one are
\be\begin{split}
	R_{Z\gamma} &\equiv \frac{\Gamma_{\eta \to Z\gamma}}{\Gamma_{\eta \to \gamma\gamma} } \approx  2 \left|\frac{A_{Z\gamma}^{\eta}}{A_{\gamma\gamma}^{\eta}} \right|^2 , \qquad
	R_{ZZ} \equiv \frac{\Gamma_{\eta \to ZZ}}{\Gamma_{\eta \to \gamma\gamma} } \approx  \left|\frac{A_{ZZ}^{\eta}}{A_{\gamma\gamma}^{\eta}} \right|^2 , \\
	R_{WW} &\equiv \frac{\Gamma_{\eta \to WW}}{\Gamma_{\eta \to \gamma\gamma} } \approx  2 \left|\frac{A_{WW}^{\eta}}{\sin^2 \theta_W A_{\gamma\gamma}^{\eta}} \right|^2 . \quad
	\label{eq:VVChannelRatio}
\end{split}\ee
The total width of such a state is of order of a few hundreds of MeV for $f \sim O(10^2 \GeV)$, therefore negligible with respect to the experimental diphoton invariant mass resolution. For a diphoton cross section of about $ 5$ fb at 13 TeV, the limits from ATLAS and CMS searches at 8 TeV on these ratios are (see e.g. Refs.~\cite{Buttazzo:2015txu,Franceschini:2015kwy})
\be
	R_{Z\gamma} \lesssim 5.6~, \qquad
	R_{ZZ} \lesssim 11~, \qquad
	R_{WW} \lesssim 36 ~.
	\label{eq:ExpLimDiboson}
\ee
As in Ref.~\cite{Redi:2016kip}, we stress that such anomaly computations are only strictly valid for $m_{\eta} \ll \Lambda$. In the case of QCD $\eta$ and $\eta^\prime$ mesons the uncertainty due to this approximation is $\sim 30\%$, and we assume a similar error also for our computations. Note that this uncertainty is still smaller than the present experimental one on the diphoton cross section.

The global symmetry breaking due to the SM gauge interactions generates, at one loop, a quadratically-divergent contribution to the mass of the pNGB which are charged under $\GG_{\SM}$, to be added in quadrature to the one due to techni-quark masses \eqref{eq:1}, which can be estimated by (see e.g. Ref.~\cite{Craig:2015lra})
\be
	\Delta m_{\pi^a}^2 \sim \frac{3 \Lambda^2}{16\pi^2} \sum_i g_i^2 C_2^i(\pi^a)~,
	\label{eq:CWpionMass}
\ee
where the sum is over the SM gauge groups under which $\pi^a$ is charged, and $C_2^i(\pi^a)$ are the corresponding Casimirs in the representation of $\pi^a$.\footnote{For the fundamental of $\SU(N)$, $C_2(N) = \frac{N^2 - 1}{2N}$, while for the adjoint $C_2(R) = N$.}
For the representations relevant to our work, this is given by
\be\begin{split}
&	\Delta m^2_{({\bf 1},{\bf 3},0)} \sim (0.13 \Lambda)^2~, \quad
	\Delta m^2_{({\bf 8},{\bf 1},0)} \sim (0.26 \Lambda)^2~, \quad
	\Delta m^2_{({\bf 8},{\bf 3},0)} \sim (0.28 \Lambda)^2~, \\
&	\Delta m^2_{({\bf 3},{\bf 1},Y)} \sim (0.17 \Lambda)^2~, \quad
	\Delta m^2_{({\bf 3},{\bf 2},Y)} \sim (0.19 \Lambda)^2~, \quad
	\Delta m^2_{({\bf 3},{\bf 3},Y)} \sim (0.21 \Lambda)^2~, \quad
	\label{eq:gauge_mass_contr}
\end{split}\ee
where we neglect the small contribution from the hypercharge in the second line, and stress that the singlet does not receive any such correction to its mass.

\subsection{Coupling to SM fermions} 

\begin{figure}[tbp]
  \centering
   \includegraphics[width=0.4\textwidth]{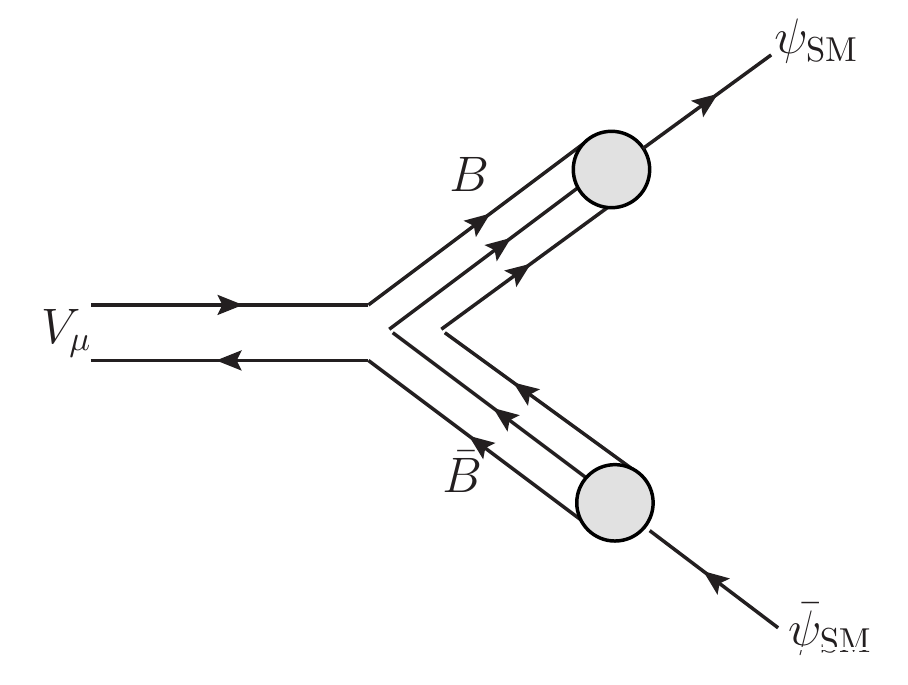}  
    \caption{ \small  Schematic diagram of the mechanism generating the coupling of the composite vector mesons $V_\mu$ with SM fermions $\psi_{\rm SM}$ via the mixing with TC baryons $B$.
        \label{Fig:VBB} }
\end{figure}

We assume that the new sector interacts with the SM fermions not only via SM
gauge interaction, but 
also via an additional (flavor) dynamics occurring at a scale $\Lambda_F > \Lambda$ and respecting an approximate 
$\U(2)^5  =  \U(2)_{q_L} \times \U(2)_{u_R} \times \U(2)_{d_R}   \times \U(2)_{\ell_L} \times \U(2)_{\ell_R} $ 
flavor symmetry.  The latter implies that only the third generation of SM fermions (singlets under $\U(2)^5$) can have non-vanishing Yukawa couplings in the limit of exact symmetry~\cite{Barbieri:2011ci}. We assume that also the techni-quarks are $\U(2)^5$ singlets, such that only third generation SM fermions can have an effective linear mixing with the composite states in the limit of unbroken flavor symmetry (thus providing a concrete realization of the mechanism proposed in Ref.~\cite{Glashow:2014iga}).

In order to construct a predictive framework, avoiding a detailed description of the dynamics occurring 
above the scale $\Lambda$, we assume that the coupling of the SM fermions with the composite sector 
is effectively well described by the mixing of the SM fermions with a proper set of composite baryons \cite{Kaplan:1991dc}.
The strong couplings of the composite baryons to the composite mesons then allow us to build 
an effective coupling of the composite  mesons (both pNGB and vector resonances) 
to SM fermion currents, as shown schematically in Fig.~\ref{Fig:VBB}.
We note that in order to have a sizable mixing of SM fermions with the TC baryons, as required by the flavor anomalies, the scale $\Lambda_F$, where the effective operators responsible for such mixing arise, should lie not too far above $\Lambda$. Nevertheless, the scale of the dynamics responsible for the flavor-symmetry-breaking and the generation of the spurions can be much higher. More details about the mixing with the TC baryons and possible UV completions are given in Sect.~\ref{sec:mixing}.

The vector mesons naturally appear in our framework as the lightest composite states above the pNGB.
Their masses are expected to be $m_\rho \sim g_\rho f$ with  $1 \ll g_\rho < 4\pi$, and thus 
somewhat lighter than the strong coupling scale $\Lambda \sim 4 \pi f$ (similarly to the $\rho$ meson in QCD). 
It is then possible to write an effective low-energy Lagrangian for these resonances. In particular, in the following we are interested in the lowest-lying spectrum of spin-1 resonances, 
i.e.~composite states of $\bar{\psi}_i \psi_j$ that belong to the same $\GG_{\SM}$ representations of the pNGB. 
As we will show, a generic model predicts colorless triplets and singlets
as in QCD, but also vector leptoquarks (LQ) and color-octets with various $\SU(2)_L$ charges. 
Depending on the specific choice of fundamental fermions, only some of these states 
will be generated and with a given multiplicity. 
Given the flavor couplings discussed above,  some of these states can be identified with
heavy vectors invoked  in Ref.~\cite{Greljo:2015mma,Barbieri:2015yvd} 
 to explain the experimental anomalies in $B$ decays.

\section{Explicit Models}
\label{sec:models}

In the following we discuss two representative choices for the $\GG_{\SM}$ representations of the techni-quarks, leading to the models that can accommodate for the diphoton excess as well as $B$ meson decay anomalies.

\begin{table}[t]
\renewcommand{\arraystretch}{1.2}
$$\begin{array}{c||  l | l | l}
\text{Model}	&\text{Flavor~structure} &  \GG_{\SM} \text{ irrep} & \text{pNGB~Mass} \\\hline
\multirow{4}{*}{\bf I}	&(\bar{Q} Q) &	({\bf 8},{\bf 1},0) &	m^2_{(\bar{Q} Q)} = 2 B_0 m_Q \\
	&(\bar{L} Q)~   + ~ {\rm h.c.}  & ({\bf 3},{\bf 2}, \Delta Y)  +  {\rm h.c.} & 	m^2_{(\bar{L} Q)} = B_0 (m_L + m_Q)  \\
	&(\bar{L} L)  & ({\bf 1},{\bf 3},0) &	m^2_{(\bar{L} L)} =  2 B_0 m_L \\
	&3 (\bar{L} L) - 2(\bar{Q} Q) & ({\bf1},{\bf1},0) &	m^2_{\eta} = \frac{2}{5} B_0 (3 m_L + 2 m_Q)  \\\hline
\multirow{4}{*}{\bf II}	&(\bar{Q} Q) &	({\bf8},{\bf3},0), ~ ({\bf8},{\bf1},0),~ ({\bf1},{\bf3},0) &	m^2_{(\bar{Q} Q)} = 2 B_0 m_Q \\
	&(\bar{L} Q)~  + ~ {\rm h.c.} 
    & ({\bf3},{\bf1}, \Delta Y), ~ ({\bf3},{\bf3}, \Delta Y)  +  {\rm h.c.} & 	m^2_{(\bar{L} Q)} = B_0(m_L + m_Q)  \\
	&(\bar{L} L) & ({\bf 1},{\bf 3},0) &	m^2_{(\bar{L} L)} = 2 B_0 m_L \\
	&3(\bar{L} L) - (\bar{Q} Q) & ({\bf1},{\bf1},0) &	m_{\eta}^2 = \frac{1}{2} B_0 ( 3 m_L + m_Q)
\end{array}$$
\caption{\label{tab:minimal} pNGB   spectrum in the minimal model (I)  and 
in the extended model (II). $\Delta Y = Y_Q - Y_L$.}
\end{table} 
 
\subsection{Minimal model: $Q = ({\bf N_{TC}}, {\bf 3}, {\bf 1}, Y_Q)$ and $L = ({\bf N_{TC}}, {\bf 1}, {\bf 2}, Y_L)$}
\label{sec:minimal}
Since our aim is to couple the composite baryonic resonances to the left-handed quark and lepton currents of the SM, in order to explain the $B$ anomalies, the strong sector has to contain both colored triplets and $\SU(2)_L$ doublets.
The minimal fermion content able to achieve the proposed goal consists of two
sets of techni-quarks, $Q = ({\bf N_{TC}}, {\bf 3}, {\bf 1}, Y_Q)$ and $L = ({\bf N_{TC}}, {\bf 1}, {\bf 2}, Y_L)$,
corresponding to 5 independent Dirac fermions from the point of view of the new strong dynamics. Interestingly, 
this set of fermions fits into a {\bf 5} + $\bar{\bf5}$ of $\SU(5)_{\rm gauge}$.

\begin{figure}[tbp]
  \centering
   \includegraphics[width=0.47\textwidth]{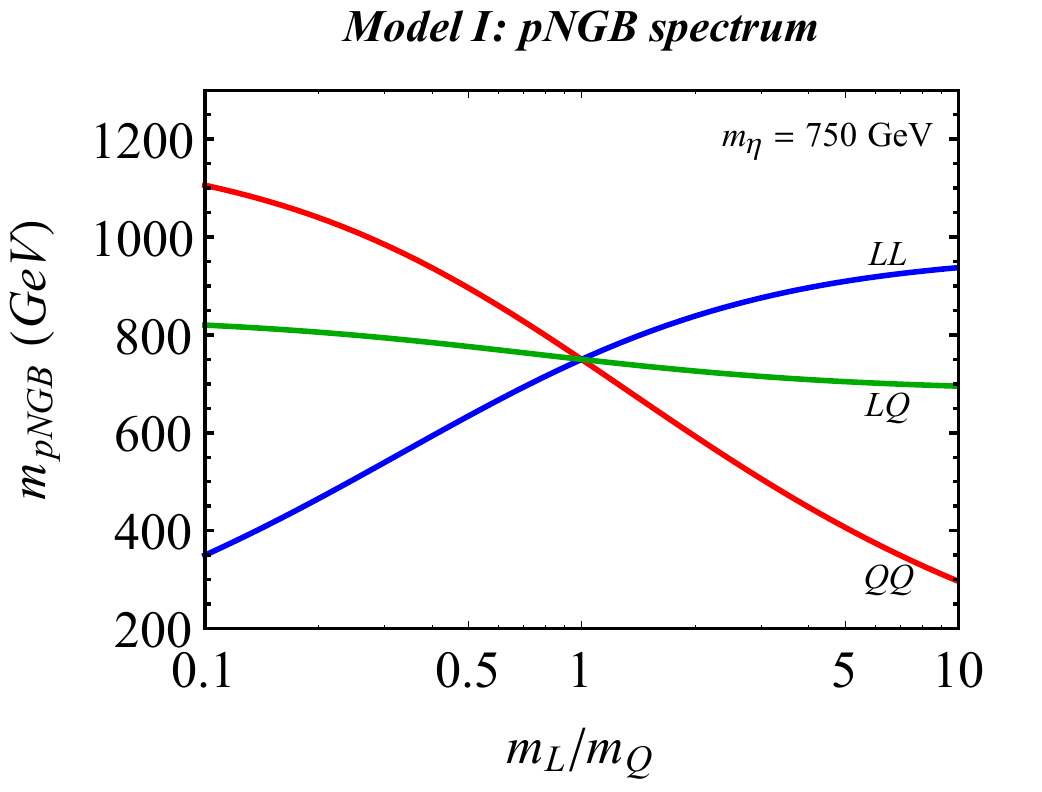}~
   \includegraphics[width=0.47\textwidth]{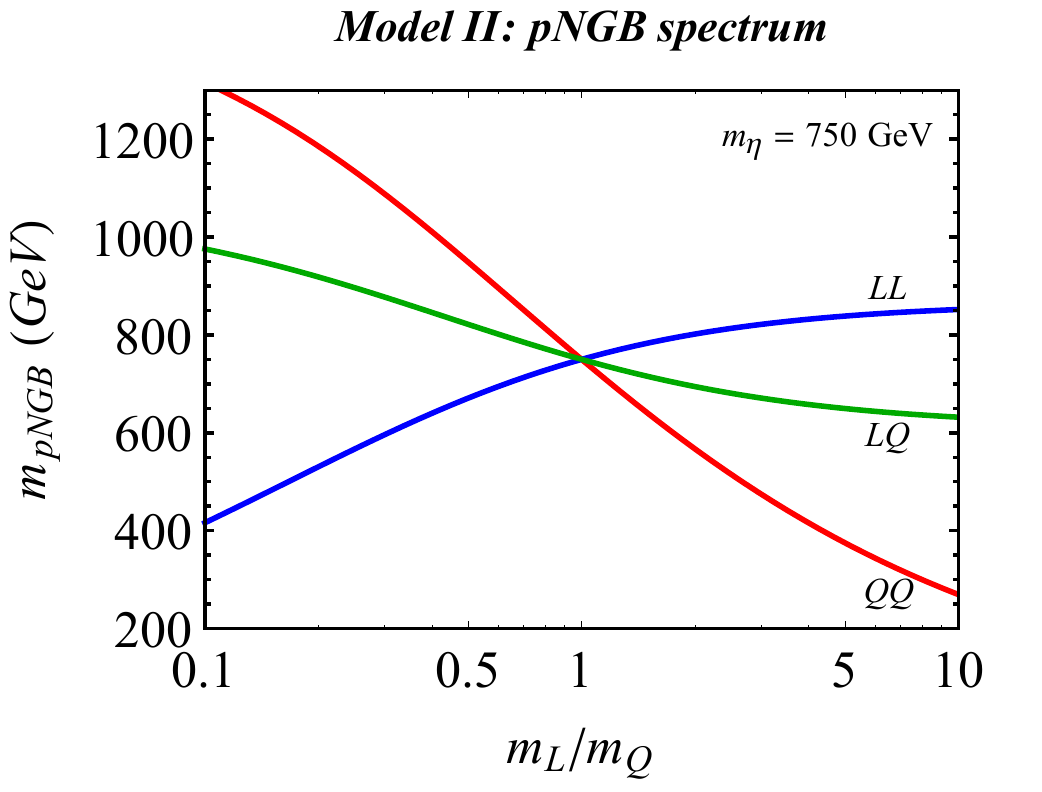}  
    \caption{ \small  pNGB spectrum in the minimal model (left) and the extended model (right), without the gauge contribution of Eq.~\eqref{eq:gauge_mass_contr}. The mass of the diphoton candidate resonance is fixed to $m_\eta = 750$~GeV.
    \label{Fig:spectrum} }
\end{figure}

The  breaking of the $\SU(5)_L \times \SU(5)_R$ global symmetry gives rise to 24 pNGB as listed in Table~\ref{tab:minimal} (up), where 
$\Delta Y = Y_Q - Y_L$. 
The anomaly couplings of the singlet ($\eta$) are given by
\be
	A_{BB}^\eta = 2 \sqrt{\frac{3}{5}} N_{TC} \left(Y_Q^2 - Y_L^2 \right)~,	\qquad 	
	A_{WW}^\eta = - \frac{1}{2} \sqrt{\frac{3}{5}} N_{TC}~, \qquad
	A_{gg}^\eta = \frac{N_{TC} }{\sqrt{15}}~.
	\label{eq:eta_VV}
\ee
In particular, we consider two choices for the hypercharges  of the TC fermions: $\{ Y_Q, Y_L \} = \{ -\frac{1}{6} , \frac{1}{6} \}$ and
$\{ Y_Q, Y_L \} = \{ 0 , -\frac{1}{6} \}$.  As we discuss below, for $N_{TC}=3$ these 
choices allow to construct techni-baryons with the quantum numbers of the SM quark and lepton doublets. 
\begin{figure}[tbp]
  \centering
   \includegraphics[width=0.55\textwidth]{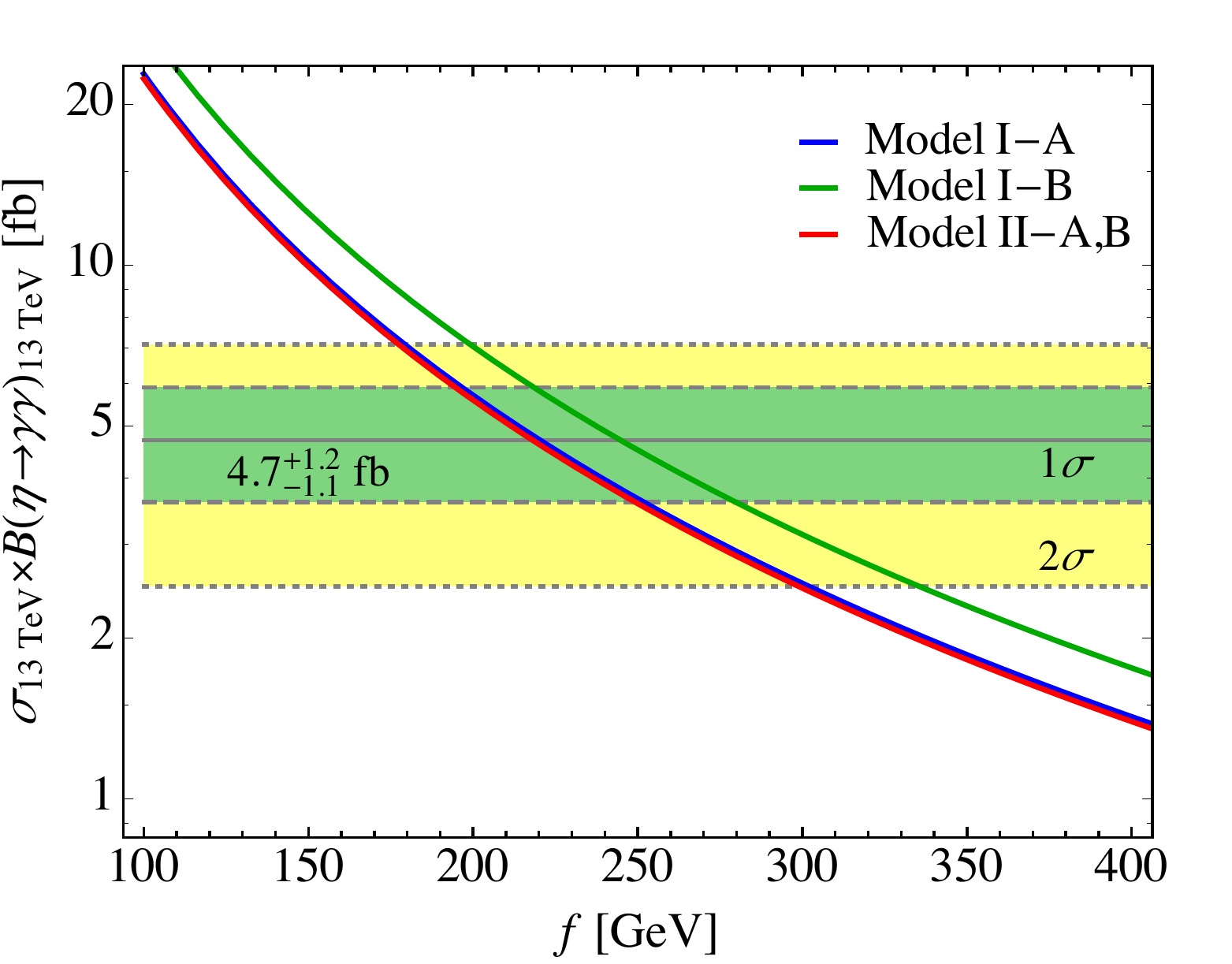}  
    \caption{ \small  Diphoton cross section from the pseudoscalar singlet $\eta$ via the anomalous coupling to photons and gluons, as a function of $f$ for $N_{TC} = 3$. The blue (green) line is for the model I with the first (second) hypercharge assignment while the red line is for the model II. In this plot we assume the $\eta$ decay width to be dominated by decays to gluons.
        \label{Fig:diphoton} }
\end{figure}
Fitting the diphoton excess, see Eq.~\eqref{eq:diphoton_comb},
and neglecting other decay channels than those in dibosons (see Sect.~\ref{sec:eta_ttbar} for a discussion on this point) we get, for the first (second) hypercharge assignment:
\be
	f \approx 71 ~(79) ~ N_{TC}~\GeV~.
	\label{eq:fmin}
\ee
This is also shown in Fig.~\ref{Fig:diphoton}.
Given the hypercharge assignments above, we can predict the ratios of the cross section in other diboson channels with respect to the diphoton one, Eq.~\eqref{eq:VVChannelRatio},
\be
\begin{array}{l}\\
\text{Model I}
\end{array}
\qquad {\renewcommand{\arraystretch}{1.1}
\begin{array}{l c | c c c}
			&	( Y_Q, Y_L )	& R_{Z\gamma} & R_{ZZ} & R_{WW} \\ \hline
	\text{A:}\quad& ( -\frac{1}{6} , \frac{1}{6} ) & 6.7 & 11 & 37 \\
	\text{B:}\quad& ( 0 , - \frac{1}{6} ) & 5.0 & 9.1 & 30
\end{array}} ~.
\ee
Both hypercharge assignments predict signals in these channels comparable
with the present experimental bounds shown in Eq.~\eqref{eq:ExpLimDiboson}.
This implies that, in this model, signals in all these diboson channels are expected to show up very soon during the forthcoming LHC run.
Setting $m_\eta=750$~GeV and varying the  ratio $m_L/m_Q$ leads to  the spectrum of pNGB masses reported in Fig.~\ref{Fig:spectrum} (left), before the gauge corrections in Eq.~\eqref{eq:gauge_mass_contr} (to be summed in quadrature).
Fixing $f \sim 220 \GeV$ and $B_0 \approx 20 f$, from $m_\eta = 750 \GeV$ we obtain that the elementary TC fermion masses are $m_L \lesssim 100$~GeV and $m_Q \lesssim 150$~GeV, therefore they are both much lighter than $\Lambda$.

The spectrum of vector mesons states, $|V_{ij} \rangle =  | (\bar{\psi}_i \psi_j)_{J=1} \rangle$,  can also be easily deduced 
from Table~\ref{tab:minimal}: it consists of a color-octet $\SU(2)_L$-singlet ($\mathcal{V}^A$), a complex 
$\SU(2)_L$-doublet leptoquark ($U_i^\alpha$), and finally one $\SU(2)_L$-triplet and two $\SU(2)_L$-singlet
colorless vectors as in QCD. In analogy with QCD, we label and normalize the colorless vectors as
\be
| \rho^a \rangle_{ ({\bf 1},{\bf 3},0)}  = \frac{1}{\sqrt{2} } | \bar L  \sigma^a  L \rangle ~, \quad 
| \omega \rangle_{ ({\bf 1},{\bf 1},0)}  = \frac{1}{\sqrt{2} } | \bar L    L \rangle  ~, \quad
| \phi \rangle_{ ({\bf 1},{\bf 1},0)}  = \frac{1}{\sqrt{3} } | \bar Q Q \rangle  ~.
 \ee
The mass of a generic vector meson $V_{ij}$ can be written as 
\be
m^2_{V_{ij}} =  c^2_0  (4\pi f)^2 + c^2_1 B_0 ( m_{\bar{\psi}_i} + m_{\psi_j} )~,
\label{eq:vr-mass}
\ee
with $c_0 < 1$ and  $c_1=O(1)$. \footnote{In the $N_{TC}=3$ case, extrapolating from the QCD spectrum, 
we expect $c^2_0 \approx 0.5$ and $c^2_1 \approx 1$.} 
 
For later purpose, we notice that for $N_{TC}=3$ we can construct two techni-baryons with 
the quantum numbers of the SM quark and lepton doublets (or anti-doublets), namely 
\be
| \bar B_\ell (B_\ell) \rangle_{ ({\bf 1},{\bf 2},  \pm 1/2) }  \propto  |   L   L   L \rangle  \qquad {\rm and} \qquad
| \bar B_q  \rangle_{ ({\bf \bar 3},{\bf 2},  - 1/6) }   \propto  |     Q    Q  L \rangle~, 
\ee
where the hypercharge assignment of the   $|  L   L   L \rangle$  state depends on the two possible 
choices for  $Y_Q$ and $Y_L$. In the case $\{ Y_Q, Y_L \} = \{ 0 , -\frac{1}{6} \}$ we can also construct 
a techni-baryon with the quantum numbers of $d_R$, namely $|QLL\rangle$,\footnote{Although the mixing of such a state with $d_R$ would violate lepton and baryon number conservation.}
while no other techni-baryon with the quantum numbers of SM fermions is allowed for
$\{ Y_Q, Y_L \} = \{ -\frac{1}{6} , \frac{1}{6} \}$.\footnote{A comment is in order about the role of the spin in baryon spectroscopy. In particular, the spin and flavor groups can be embedded in $\SU(10)\supset \SU(5)_F \times \SU(2)_S$ and the baryons are formed as: ${\bf 10}\times {\bf 10}\times {\bf 10} = {\bf 120} + {\bf 220} + 2 ({\bf 330})$. The technicolor wave function is antisymmetric (for $N_{TC}=3$), since hadrons are $\SU(N_{TC})$ singlets. Therefore, baryons have to occupy the $\bf{220}$ of $\SU(10)$, which is symmetric.
Under $\SU(5)_F \times \SU(2)_S$, ${\bf 220} = (\bar {\bf 40},{\bf 2})+( \bar {\bf 35},{\bf  4})$, therefore
spin 1/2 baryons belong to the $ \bar{\bf 40}$ of $\SU(5)_F$, which decomposes as $\bar{\bf 40} = ({\bf 1},{\bf 2})+({\bf 3},{\bf 1})+({\bf \bar{3}},{\bf 2})+({\bf 3},{\bf 3})+({\bf 6},{\bf 2})+({\bf 8},{\bf 1})$ under $\SU(3)_c\times\SU(2)_L$.}

There are two independent, unbroken, and non-anomalous $\U(1)$ factors: the diagonal subgroup of $\SU(5)_V$, and $\U(1)_V$. These two global charges also correspond to the $Q$- and $L$-numbers, and after mixing of the techni-baryons with the SM quark and lepton doublets, are matched to the SM baryon and lepton numbers.
One combination of them is identified with the SM hypercharge. Notice that, since the hypercharge assignment of the techni-quarks is never such that it can be matched to the diagonal $\SU(5)$ generator, an embedding of the model into $\SU(5)_{\rm gauge}$ requires also the second $\U(1)$ to be gauged.

\subsection{Extendend model: $Q = ({\bf N_{TC}}, {\bf 3}, {\bf 2}, Y_Q)$ and $L = ({\bf N_{TC}}, {\bf 1}, {\bf 2}, Y_L)$}
\label{sec:leptoquark}

An extended spectrum of vector-meson resonances, including in particular vector leptoquarks, is obtained 
if we assume that $Q$ is a doublet of $\SU(2)_L$, rather than a singlet as in the minimal model, $Q = ({\bf N_{TC}}, {\bf 3}, {\bf 2}, Y_Q)$ and $L = ({\bf N_{TC}}, {\bf 1}, {\bf 2}, Y_L)$. 
This leads to a $\SU(8)_L \times \SU(8)_R$ global flavor symmetry.
The corresponding set of 63 pNGB is listed in Table~\ref{tab:minimal} (bottom). 
The anomalous couplings of the singlet pseudoscalar meson are
\be
	A_{BB}^\eta = \sqrt{3} N_{TC} \left( Y_Q^2 - Y_L^2 \right)~,	\qquad 	
	A_{WW}^\eta = 0~, \qquad
	A_{gg}^\eta = \frac{N_{TC}}{4\sqrt{3}}~.
	\label{eq:eta_VV_Mod2}
\ee
The hypercharge assignments that allow to construct techni-baryons with the quantum numbers of SM quark  and lepton doublets, for $N_{TC}=3$, are 
\be\begin{array}{c | c c c c c}\label{modelII}
Y_Q & 1/6 & 1/2 &  1/6 & 0 & -1/6 \\ \hline
Y_L & -1/2 & -1/6 &  0 & -1/6 & 1/6
\end{array}~.
\ee 
The latter leads to a vanishing anomalous coupling to photons in Eq.~(\ref{eq:eta_VV_Mod2}), therefore we discard that option. In the first two (third and fourth) cases the diphoton excess can be fit with
\be
	f \approx 70 ~ (9)~ N_{TC}\GeV~,
	\label{eq:fnonmim}
\ee
which shows that only the first two are phenomenologically viable for $N_{TC} = 3$.
The prediction for the cross sections in the other diboson channels at 750 GeV are given by
\be
\begin{array}{l}\\
\text{Model II}
\end{array}
\qquad
{\renewcommand{\arraystretch}{1.1}\begin{array}{l c | c c c}
			&	( Y_Q, Y_L ) & R_{Z\gamma} & R_{ZZ} & R_{WW}\\ \hline
	\text{A:}\quad& (\frac{1}{2}, -\frac{1}{6}) & \multirow{2}{*}{0.6} & \multirow{2}{*}{0.09} & \multirow{2}{*}{0}\\
	\text{B:}\quad& (\frac{1}{6}, -\frac{1}{2})
\end{array}}~.
\ee
In Fig.~\ref{Fig:spectrum} (right) we show the pNGB spectrum as a function of $m_L/m_Q$ fixing $m_\eta = 750 \GeV$, again neglecting the gauge corrections of Eq.~\eqref{eq:gauge_mass_contr}.
In this case the TC quark masses are bounded by $m_L \lesssim 100 \GeV$ and $m_Q \lesssim 300 \GeV$, again much smaller than $\Lambda$.

As anticipated, the interest of considering this extended model lies  in the modified spectrum of vector mesons, that  contains more exotic states. In addition to the color-octet $\SU(2)_L$-singlet  ($\mathcal{V}^A$) there is also a color-octet $\SU(2)_L$-triplet ($\mathcal{V}^{A,a}$), and vector leptoquarks appears both in singlet ($U$)  and triplet ($\mathcal{U}^a$) representations of $\SU(2)_L$. The colorless states include the $\rho^a$,  $\omega$, and $\phi$  appearing already in the minimal model, plus a second $\SU(2)_L$ triplet 
\be
| (\rho^\prime)^a \rangle_{ ({\bf 1},{\bf 3},0)}   \propto  | \bar Q  \sigma^a  Q \rangle~.
\ee

The only  techni-baryons with the quantum numbers of  SM fermions, for $N_{TC}=3$ and for the first two hypercharge assignments of \eqref{modelII}, are~\footnote{With a proper combination of spin and $\SU(2)_L$ indices it is possible to build a completely antisymmetric wave function for  the $|QQQ \rangle$ baryon that is a doublet
of spin and $\SU(2)_L$,  
and a singlet of color and TC indices.} 
\begin{align}
  \text{A:}& \qquad | B_\ell  \rangle_{ ({\bf 1},{\bf 2},  -1/2) }  \propto  |   LLL \rangle
  \qquad\text{and}\qquad
  | B_q  \rangle_{ ({\bf 3},{\bf 2},    1/6) }   \propto  |   Q LL \rangle\,,\\
  \text{B:}& \qquad | \bar B_\ell  \rangle_{ ({\bf 1},{\bf 2},  1/2) }  \propto  |   QQQ \rangle
  \qquad\text{ and}
  \qquad
  | \bar B_q  \rangle_{ ({\bf \bar 3},{\bf 2},    -1/6) }   \propto  |   Q QL \rangle\,.
\end{align}

\section{Couplings of the composite mesons to SM fermions}
\label{sec:Flavor-mixing}

Most of the heavy vectors appearing in the composite models could have effective couplings of the type 
$ V_{A}^\mu J_{A,{\rm SM}}^\mu$, where $J_{A,{\rm SM}}^\mu$ is an appropriate current constructed from SM fermions. 
In general, several of such couplings are severely  constrained from 
flavor-physics experiments and collider searches. However, as shown in 
Ref.~\cite{Greljo:2015mma,Barbieri:2015yvd} (see also \cite{Glashow:2014iga}), 
a framework that is naturally consistent with existing bounds and could possibly alleviate the present tensions in $B$ physics is obtained under the assumption that the heavy vectors couple mainly to third-generation fermions.
More explicitly, in order to obtain a maximally predictive framework, in the following we assume that 
\begin{itemize}
\item[I.] The interactions between the heavy vectors and the SM fermions arise only as the effect of the 
mixing between SM fermions and techni-baryons (that, in turn, are strongly coupled to the heavy vectors).
\item[II.] The above mixing respects an approximate $\U(2)^5$ flavor symmetry, under which techni-quarks 
and third-generation SM fermions are singlets.
\item[III.]  The leading corrections to the exact $\U(2)^5$ limit are 
obtained introducing two spurions transforming as doublets of 
 $\U(2)_{q_L}$ and $\U(2)_{\ell_L}$, respectively (consistently with the dominant   $\U(2)^5$  breaking terms 
 observed in the SM quark Yukawa couplings~\cite{Barbieri:2011ci,Barbieri:2012uh}).
\end{itemize}
As shown in Sect.~\ref{sec:models}, with a proper choice of hypercharge assignments we can restrict the 
attention to models where the only techni-baryon with the quantum numbers of SM fermions 
are the $B_q$ and $B_\ell$ states (that could mix with $q_L$ and $\ell_L$, respectively).
Under this further assumption, the flavor structure of the vector resonances is particularly simple,
as discussed below. Possible consequences of  less restrictive assumptions about the 
underlying mechanism responsible for the mixing between TC and SM fermions will be briefly 
outlined at the end of this section.

\subsection{Mixing with TC baryons}
\label{sec:mixing}

On general grounds, we can parameterize the vectorlike interactions 
of the vector mesons with the TC baryons as follows
\ba
\cL_U &=& g_\rho~a^U_{q \ell} ~ \bar B_q \gamma^\mu B_\ell ~U_\mu+\rm{h.c.}~,\\
\cL_\omega &=& g_\rho ~ (a^\omega_q ~\bar B_q \gamma^\mu B_q+a^\omega_\ell~\bar B_\ell \gamma^\mu B_\ell) ~ \omega_\mu~,
\ea
and similarly for the $\SU(2)_L$ triplets $\rho_\mu^a$ and $\mathcal{U}^a_\mu$ with the corresponding 
baryonic currents ($\bar B_{q(\ell)} \gamma^\mu \sigma^a B_{q(\ell)}$).
The parameters $a^V_{x}$ depend on the details of the techni-dynamics and are difficult to be estimated. However, 
for the case $N_{TC}=3$, we expect the following $O(1)$  
relations to hold:
\begin{itemize}
\item[ ($L\bar L$)] The valence techni-quarks of the $\rho$ and $\omega$ states
are such that they can couple via connected diagrams to baryons containing an $L$ techni-quark. In the minimal model, as well as in model II\,A, they could couple to both $B_q$ and $B_\ell$, and we expect 
both $a^{\rho,\omega}_{q}$ and $a^{\rho,\omega}_{\ell}$ to be $O(1)$. In model II\,B they can couple only to $B_q$, and we thus expect $a^{\rho,\omega}_{\ell}\ll 1$.
\item[ ($Q\bar Q$)] The valence techni-quarks of the $\phi$, $\mathcal{V}$ and 
$\rho^\prime$ states (some of them only present in the extended model) are such that they could couple only to 
baryons containing a $Q$ field via connected diagrams. In the models I and II\,A we thus expect $a^{\phi, \mathcal{V}, \rho^\prime}_{q} = O(1)$ and $|a^{\phi, \rho^\prime}_{\ell}| \ll1$, while in model II\,B all the coefficients are expected to be of $O(1)$, excluding $a^{\mathcal{V}}_{\ell}$ which is always zero by gauge symmetry.\footnote{In the minimal model, a na\"ive evaluation of the relevant connected amplitudes leads to the relations $2 a^{\omega}_q=a^{\rho}_{q} = a_q^\mathcal{V}
= \sqrt{3/2} a^{\phi}_q$  and $a_\ell^\mathcal{V} = a^{\phi}_\ell =0$.}
\item[ ($L\bar Q$)] As far as leptoquarks are concerned, in the minimal model they do not have quantum numbers that
allow a coupling to $B_q$ and $B_\ell$ alone; in the extended model we expect both 
$a^{U}_{ q \ell}$ and $a^{\mathcal {U}}_{ q \ell}$ to be $O(1)$. 
\end{itemize}

\medskip
\noindent
For our purposes it is not necessary to specify the underlying dynamics responsible for  
the mixing between TC baryons and SM fermions. However, we note that a simple UV completion can be obtained via the exchange of a scalar field with mass 
$m_F\equiv \Lambda_F$ (around or above 
the TC scale $\Lambda$)\footnote{In order to have a sizable mixing, either  $\Lambda_F \sim \Lambda$ or, 
if $\Lambda_F \gg \Lambda$, the four-fermion operator should have a sizable anomalous dimension.
The latter option has recently been shown to be difficult to achieve in realistic models~\cite{Pica:2016rmv}. Still, it is worth to note that other four-fermion operators made of SM fields, which could mediate potentially dangerous flavor effects, cannot have such a sizable anomalous dimension since they are necessarily formed by conserved currents.}
charged under $\SU(N_{TC})\times \GG_{\SM}$. A detailed realization of a very similar setup was proposed long ago by Kaplan 
in Ref.~\cite{Kaplan:1991dc}. Following this work we can introduce  extra 
elementary scalars in the fundamental of $\SU(N_{TC})$ and also charged under the SM gauge group, with Yukawa-type interaction to SM fermions and TC quarks. 
As an explicit example, let us take the minimal model (version A) and let us introduce two complex scalar fields
\be
\phi \equiv ({\bf \bar 3},{\bf 1},1)_{1/3}~,\;\chi\equiv ({\bf \bar 3 },{\bf \bar 3 },1)_{-1/3}~,
\ee
with the following Yukawa-type interaction
\be
\mathcal L \supset y_\phi \bar \ell^C_L L_L \phi + z_\phi \bar L^C_L L_L \phi^* + y_\chi \bar q_L^C L_L \chi + z_\chi \bar Q^C_L Q_L \chi^*+\rm{h.c.}~.
\ee
Provided that $m_{\phi,\chi} \gsim \Lambda$, the scalars can be integrated out obtaining the following dimension-6 operators
\be
\mathcal{L}_{\rm eff} \supset - \frac{y_\phi z_\phi}{m_\phi^2}~ \bar \ell^C_L L_L \bar L^C_L L_L - \frac{y_\chi z_\chi}{m_\chi^2}~\bar q_L^C L_L \bar Q^C_L Q_L~,
\ee 
that generate the desired mixing.

As a result of the effective couplings between SM fermions and TC baryons we can write
\be
B_q \to \kappa_q \chi^q_i q_L^i~,~B_\ell \to \kappa_\ell \chi^\ell_i \ell_L^i~,
\label{eq:Btoq}
\ee
where the coefficients $\chi^{q(\ell)}$ can be decomposed as follows (in flavor space):
\be
\chi^{q(\ell)}=\left(\begin{array}{c}
\varepsilon_{1}^{q(\ell)}\\
\varepsilon_{2}^{q(\ell)}\\
1
\end{array}\right)~.
\ee
Here $\varepsilon_{1}^{q(\ell)} \ll \varepsilon_{2}^{q(\ell)} \ll 1$ denote the two components of the 
$\U(2)_{q_L(\ell_L)}$ breaking spurions, and the flavor basis is defined as the $\SU(2)_L$-invariant 
basis where down-type quarks are diagonal.
In the quark sector  we expect the spurion to 
be aligned with the CKM entries describing the mixing between third and light generations, 
namely   $(\varepsilon^q_1, \varepsilon^q_2) = \xi \times (V_{t d}, ~V_{t s})$. The parameter $\xi$, 
expected to be of $O(1)$, defines the orientation of the $\U(2)^5$ singlets 
with respects to the quark mass eigenstates. The limit $\xi \to 0$ corresponds to a perfect 
alignment of the $\U(2)^5$ singlets to the third generation of SM quarks 
in the  down-type mass-eigenstate basis. 

Using the relations (\ref{eq:Btoq}) we get
\ba
\cL_U &=& g_\rho~a^U_{q \ell}~\kappa_q \kappa_\ell~\beta_{i j}~ \bar q_L^i \gamma^\mu \ell^j_L ~U_\mu+\rm{h.c.}~,\\
\cL_\omega &=& g_\rho ~ (a^\omega_q~\kappa_q^2~\lambda_{i j}^q~\bar q_L^i \gamma^\mu q_L^j+a^\omega_\ell~\kappa_\ell^2~\lambda_{i j}^\ell~\bar \ell_L^i \gamma^\mu \ell_L^j) ~ \omega_\mu~\label{eq:vect-fl},
\ea
where
\be
\lambda^{q(\ell)}_{i j}=\chi^{q(\ell)*}_i \chi^{q(\ell)}_j~, \qquad ~\beta_{i j}=\chi^{q*}_i \chi^{\ell}_j~.
\ee
In the flavor fit presented in Sect.~\ref{sec:flavor} we treat as independent parameters 
\be
\lambda_{b s}^q \quad{\rm and}\quad \lambda^{\ell}_{\tau \mu}~,
\ee
in terms of which the other flavor couplings appearing in Eq.~(\ref{eq:vect-fl}) 
assume the form
\be
\lambda^{\ell}_{\mu \mu} = |\lambda_{\tau \mu}^\ell|^2~,  \quad
~\beta_{s \mu} = \lambda_{b s}^{q*}\lambda_{\tau \mu}^\ell~,  \quad
~\beta_{s \tau} =\lambda_{b s}^{q*} ~, \quad
~\beta_{b \mu} = \lambda_{\tau \mu}^\ell~.
\label{eq:minflav}
\ee

\subsection{Decay of the singlet pNGB to $t \bar{t}$}
\label{sec:eta_ttbar}

A possible decay channel of the pNGBs in our model is to SM fermions, arising from meson-baryon interactions and mixing of the latter with SM fermion doublets, as discussed above and through similar diagrams as in Fig.~\ref{Fig:VBB}. In the case of the singlet $\eta$ responsible for the diphoton excess, such a decay could in principle decrease the signal rate by increasing the total decay width, thus requiring too small values of $f$ to fit the signal.
Contrary to the couplings of the vector mesons to SM fermions, the pNGB couplings have a further mass suppression. The only relevant couplings are thus the one proportional to the top mass.

In order to estimate such decays, we parametrize the interaction of $\eta$ with baryons as
\be
	\Delta \cL = i g_{\eta B B} \eta ( \bar{B}_\ell \gamma^5 B_\ell + \bar{B}_q \gamma^5 B_q).
	\label{eq:eta_bar_Lagr}
\ee  
After mixing of the TC baryons with the SM doublets this interaction generates
\be
	\Delta \cL = i \frac{g_{\eta B B}}{m_B} \eta \left( \kappa_\ell^2 m_\tau \bar{\tau} \gamma^5 \tau + \kappa_q^2 m_t \bar{t} \gamma^5 t + \kappa_q^2 m_b \bar{b} \gamma^5 b \right)~.
	\label{eq:eta_ttbar_Lagr}
\ee
Given the flavor composition of $\eta$, we expect  a significant suppression of $g_{\eta B B}$ compared to $g_\rho$. In particular, a na\"ive  counting of the relevant connected amplitudes leads to $g_{\eta B B}  \sim g_\rho/\sqrt{15}$ for the minimal model (I).  
The decay width of the pseudo-scalar in $t \bar{t}$ is given by 
\be 
\Gamma_{tt} \approx (3.3~{\rm GeV}) \times  \frac{m^2_\eta}{m^2_B} \left| g_{\eta B B} \kappa_q^2  \right|^2.
\ee
Including this decay channel, for $m_\eta = 750 \GeV$, $N_{TC} = 3$, $m_B \sim 4 \pi f$, $g_q = 2 g_\rho \kappa_q^2  \approx 4$, and $g_{\eta B B} \approx g_\rho/4$, in model I\,A\,(B) the diphoton signal can be fit with $f \approx 190 \GeV$ ($210 \GeV$). This corresponds to $\mathcal{B}(\eta \to t\bar{t}) \approx 19\%$, thus justifying a posteriori the approximation of having neglected this decay channel when fitting the diphoton excess in Sect.~\ref{sec:models}.

\subsection{Non-mixing operators}
\label{sec:non-mixing}

The assumption that the effective couplings of quarks and leptons to the resonances are entirely mediated by the mixing of SM fermions and TC baryons is a strong dynamical one.
As shown above, it leads to a very predictive framework; however, it is not the only viable option.
In general, we cannot exclude that the UV dynamics responsible for this mixing induces also 
(or even mainly) additional structures. For instance, an interesting possibility is that of flavor mediators 
generating effective four-fermion interactions of the type
\be
\frac{c_f^{ij}}{\Lambda^2_F} (\bar \psi_{\rm TC} \gamma_\mu \psi_{\rm TC}) ( \bar f^i_{\rm SM} \gamma_\mu f^j_{\rm SM})~,
\label{eq:nonminflav}
\ee
where $\psi_{\rm TC}=L,Q$ and $f_{\SM}=q_L,\ell_L$. 
With structures of this type, and introducing additional $\U(2)^5$  
breaking spurions,  there is clearly more freedom in varying the effective flavor coupling 
of a given set of resonances to SM fermions. In particular, 
assuming $c_q^{ij}$ (or $c_\ell^{ij}$) transforms as a {\bf 3} of $\U(2)_{q_L}$ (or $\U(2)_{\ell_L})$ leads to violations of the
relations in Eq.~(\ref{eq:minflav}). 

A detailed analysis of these additional couplings, as well as of the underlying dynamics responsible 
for the effective mixing in Eq.~(\ref{eq:Btoq}),  is beyond the scope of this work. 
We simply note here that breaking the relations in Eq.~(\ref{eq:minflav}) leads to a
fit of low-energy data that requires a smaller fine-tuning among hadronic parameters
(see Section~\ref{sec:flavor}).

\section{Low-energy four-fermion interactions}
\label{sec:flavor}

In this section we present both the structure of four-fermion interactions obtained integrating 
out the vector resonances (at the tree-level) and 
the results of a global fit to low-energy observables. The latter is a 
generalization of the fit performed in Ref.~\cite{Greljo:2015mma}, to which we refer for 
a more extensive discussion about the explicit expressions of the 
observables and their experimental determination.\footnote{Related EFT-type 
analyses of these constraints have recently been presented in 
Refs.~\cite{Bhattacharya:2014wla,Alonso:2015sja,Freytsis:2015qca} (pure EFT approach) 
and in Refs.~\cite{Gripaios:2014tna,Calibbi:2015kma,Fajfer:2015ycq,Bauer:2015knc} (in the context of LQ models).}

\subsection{Effective interactions due to $\rho$ and $\omega$ exchange}
\label{sec:flavor1}

The tree-level exchange of $\rho$ and $\omega$ leads to the following 
non-standard four-fermion interactions~\cite{Greljo:2015mma}
\be
  \Delta \cLT_{4f} =  - \frac{1}{2 m_\rho^2} J_\mu^a J_\mu^a~,\; \; \;  \Delta \cLS_{4f} =  - \frac{1}{2 m_\omega^2} J_\mu^0 J_\mu^0~,
  \label{eq:DeltaL4f}
 \ee
 where  $J_\mu^a$ ($J_\mu^0$) is a fermion current  transforming as a $\SU(2)_L$ triplet (singlet), built in terms of the SM quarks and lepton doublets:
\ba
	J^a_\mu &=&  \gq \llq_{ij} \left(\bar q_L^i \gamma_\mu \tau^a q_L^j\right) 
+ \gl \lle_{ij} \left(\bar \ell_L^i \gamma_\mu \tau^a \ell_L^j\right)~,\\
	J_\mu^0 &=&\frac{1}{2} \gqs \llq_{ij} \left(\bar q_L^i \gamma_\mu q_L^j\right) 
+ \frac{1}{2} \gls \lle_{ij} \left(\bar \ell_L^i \gamma_\mu \ell_L^j\right)~.
\label{eq:LVf}
\ea
Here $\lambda^{q,\ell}$ are the Hermitian flavor matrices ($\lambda^{q,\ell}_{33}=1$), $\tau^a=\sigma^a/2$, and, in first approximation, in the following we assume degenerate masses $m_\rho = m_\omega$ (see Eq.~\eqref{eq:vr-mass}).
Comparing with Eq.~\eqref{eq:vect-fl}, we obtain $g_q^0 = 2 g_\rho a^{\omega}_q \kappa_q^2$, and similarly for other couplings.
Among the four-fermion operators generated by the model, the ones most relevant to flavor phenomenology are:
{\small
\ba
\Delta  \cL_{\rm c.c.}  &=&  - \frac{ \gq \gl }{2 m_\rho^2} \left[ (V \llq )_{ij}  \lle_{ab}
\left(\bar u_L^i  \gamma_\mu  d_L^j\right)    \left(\bar \ell_L^a \gamma_\mu   \nu_L^b\right)   +{\rm h.c.} \right],\\
 \Delta \cL_{\rm FCNC}  &=& 
 - \frac{ \lle_{ab}  \llq_{ij} }{4 m_\rho^2}   \left(\bar d_L^i  \gamma_\mu  d_L^j\right)  \left[  (\gqs \gls +\gq \gl) \left( \bar \ell_L^a \gamma_\mu   \ell_L^b\right)  + (\gqs \gls -\gq \gl) \left( \bar \nu_L^a \gamma_\mu   \nu_L^b\right)   \right]  \no \\
 &&  -\frac{ \lle_{ab}  (V  \llq V^\dagger )_{ij} }{4 m_\rho^2}  \left(\bar u_L^i  \gamma_\mu  u_L^j\right) \left[  (\gqs \gls -\gq \gl) \left( \bar \ell_L^a \gamma_\mu   \ell_L^b\right)  + (\gqs \gls +\gq \gl) \left( \bar \nu_L^a \gamma_\mu   \nu_L^b\right)   \right], \label{eq-eff-lag}\\
 \Delta \cL_{\Delta F=2} &=&  - \frac{ \gq^2 +(\gqs)^2 }{8 m_\rho^2}  \left[
 (\llq_{ij})^2  \left(\bar d_L^i  \gamma_\mu  d_L^j\right)^2  + 
(V \llq V^\dagger )^2_{ij}  \left(\bar u_L^i  \gamma_\mu  u_L^j\right)^2  \right], \label{eq-eff-lag-mix}\\
 \Delta \cL_{\rm LFV} &=&  - \frac{ \gl^2 + (\gls)^2  }{8 m_\rho^2}   \lle_{ab}    \lle_{cd}   
(  \bar \ell_L^a \gamma_\mu   \ell_L^b  ) ( \bar \ell_L^c \gamma_\mu   \ell_L^d )\,,\\
 \Delta \cL_{\rm LFU}  &=&   - \frac{ 1  }{2 m_\rho^2}   \left[\frac{(\gls)^2-\gl^2}{2} \lle_{ab}    \lle_{cd} + \gl^2 \lle_{ad}    \lle_{cb} \right ]   
(  \bar \ell_L^a \gamma_\mu   \ell_L^b  ) ( \bar \nu_L^c \gamma_\mu   \nu_L^d )\,. 
\ea
}

The low-energy observables entering the fit depend on the three flavor-non-universal couplings 
$\llq_{bs}$, $\lle_{\mu\mu}$, $\lle_{\tau\mu}$, and the four flavor-independent combinations 
\be
	\epsilon_{\ell,q}^{(0)} \equiv \frac{g_{\ell,q}^{(0)} \, m_W}{g \, m_\rho} \approx g_{\ell,q}^{(0)} \frac{122 \, \GeV}{m_\rho}~,
\ee
which we assume to be bounded as $|\epsilon_{\ell,q}^{(0)}| < 1$ (for $m_\rho=1.5$~TeV this implies $|g_{\ell,q}^{(0)}| \lesssim 12$).
The contributions to the flavor observables defined in Ref.~\cite{Greljo:2015mma} are summarized in 
Table~\ref{tab:FlavorFit}.

{\small\begin{table}[t]
\renewcommand{\arraystretch}{1.35}
\begin{tabular}{c|c|c}
	Obs. $\mathcal{O}_i$ & Prediction $\mathcal{O}_i(x_\alpha)$ & Experimental value \\ \hline
	$R_0 $ & $\epsilon_\ell \epsilon_q$ & $0.13 \pm 0.03$\\
	$\Delta C_9^\mu$ & $- (\pi/\alpha_{\rm em})\, \lle_{\mu\mu} (\epsilon_\ell \epsilon_q+\epsilon_\ell^{0} \epsilon_q^{0})\,  \llq_{bs}/  |V_{tb}^* V_{ts}| $ & $-0.58 \pm 0.16$\\
	$\Delta R_{B_s}^{\Delta F = 2}$ & $\big(\epsilon_q^2+(\epsilon_q^0)^2\big)  |\llq_{bs}|^2\, (  |V_{tb}^* V_{ts}|^2 R_{\text{SM}}^\text{loop})^{-1} $ & $-0.10 \pm 0.07$\\
	$\Delta R_{b\rightarrow c}^{\mu e}$ &  $2 \epsilon_\ell \epsilon_q \lle_{\mu\mu}$ & $0.00 \pm 0.01$\\
	$ R_{\tau \rightarrow \mu / e}$ & $\left| 1+\epsilon_\ell^2 \lle_{\mu \mu} + \frac{1}{2} \big((\epsilon_\ell^{0} )^2-\epsilon_\ell^2\big) | \lle_{\tau \mu} |^2 \right|^2 + \left| \frac{1}{2} \big(\epsilon_\ell^2 +(\epsilon_\ell^0)^2\big)  \lle_{\tau \mu} \right|^2$ & $1.0040 \pm 0.0032$\\
	$\Lambda_{\tau\mu}^{-2}$ & $ (G_F/\sqrt{2}) \big(\epsilon_\ell^2 +(\epsilon_\ell^0)^2\big) \lle_{\mu \mu} \lle_{\tau \mu}$ & $(0.0 \pm 4.1) \times 10^{-9\phantom{0}} \GeV^{-2}$\\
	$\Lambda_{uc}^{-2}$ & $ (G_F/\sqrt{2}) \big(\epsilon_q^2+(\epsilon_q^0)^2\big)   | V_{ub} V_{cb}^* |^2$ & $ (0.0 \pm 5.6) \times 10^{-14} \GeV^{-2}$\\
	$R_{K^{(*)}\nu}$ & $\left[2+\left|1+ (\pi/\alpha_{\rm em}) (\epsilon_\ell \epsilon_q-\epsilon_\ell^{0} \epsilon_q^{0})  \llq_{bs}/  (|V_{tb}^* V_{ts}| C_{\nu}^{\rm{SM}}) \right|^2\right]/3$ & $0.0 \pm 2.6$ \\
 \end{tabular}
 \caption{\label{tab:FlavorFit}  Observables entering in the fit, together with their predictions in terms of the model parameters describing the contributions due to 
  $\rho$ and $\omega$ exchange, and the corresponding experimental bounds (assuming Gaussian uncertainties).}
\end{table}}

\subsection{Vector color-octet contributions to $\Delta F=2$}
\label{sec:color-octet-flav}
For the color-octet singlet and triplet fields, 
\be
\mathcal{V}^A_\mu : ({\bf 8},{\bf 1},0) \qquad{\rm and}\qquad  \mathcal{V}^{A,a}_\mu : ({\bf 8},{\bf 3},0)~,
\ee 
we write the effective couplings to quark currents as follows
\be
\mathcal{L} \supset \frac{1}{2} g_O \lambda^{q}_{i j} ~ \bar q_{L}^i \gamma^\mu T^A q_{L}^j ~ \mathcal{V}^A_\mu +  g_O' \lambda^{q}_{i j} ~ \bar q_{L}^i \gamma^\mu T^A \tau^a q_{L}^j ~ \mathcal{V}^{A,a}_\mu~,
\ee
where $T^A$ are the generators of color ($T^A=\lambda^A/2$ where $\lambda^A$ are Gell-Mann matrices) and $g_O$ ($g_O'$) is the corresponding coupling constant for the color-octet, electroweak singlet (triplet), vector resonance. The singlet contribution to meson mixing is the same as in Table~\ref{tab:FlavorFit} with the replacement $(\epsilon_q^0)^2 \to - 2 \epsilon_{O}^2/3$, where
\be
	\epsilon_{O} \equiv \frac{g_{O} \, m_W}{g \, m_{O}} \approx g_{O} \frac{122 \, \GeV}{m_{O}}~.
\ee
Note that this implies a cancellation between $\phi$ and  $\mathcal{V}^A$ contributions 
to  meson mixing in the limit  $a_q^\mathcal{V} = \sqrt{3/2} ~ a^{\phi}_q$, that is what is expected by 
a na\"ive evaluation of the relevant connected TC amplitudes contributing to these couplings.

Similarly,  the contribution of the triplet is obtained via the replacement
$(\epsilon_q)^2 \to - 2 (\epsilon_{O}')^2/3$.

\subsection{Vector leptoquark contributions to semi-leptonic operators}

For the vector LQ representations
\be
\mathcal{U}^a_\mu : ({\bf 3},{\bf 3},2/3) \qquad{\rm and}\qquad  U_\mu : ({\bf 3},{\bf 1},2/3)~,
\ee 
we define the interactions to the SM quarks and leptons as follows
\ba
\cL_{\mathcal{U}} &=& g_T ~ \beta_{i j}~ \bar q_L^i \gamma^\mu \tau^a \ell_L^j ~\mathcal{U}^a_\mu +\rm{h.c.}~\label{eq:singlet-qq},\\
\cL_U &=& \frac{1}{2} g_S ~ \beta_{i j}~ \bar q_L^i \gamma^\mu \ell_L^j ~U_\mu+\rm{h.c.}~,
\label{eq:octet-qq}
\ea
where $\beta_{i j}$ are the flavor matrices ($\beta_{3 3}=1$). Matching to the low-energy effective 
theory~\cite{Dorsner:2016wpm} we find
{\small
\ba
\Delta  \cL_{\rm c.c.}  &=&  \frac{ g_T^2-g_S^2 }{4 m_{U}^2} \left[ (V \beta )_{i b}  \beta^*_{j a}
\Big(\bar u_L^i  \gamma_\mu  d_L^j\Big)    \Big(\bar \ell_L^a \gamma_\mu   \nu_L^b\Big)   +{\rm h.c.} \right],  \label{eq:lq-charged}\\
 \Delta \cL_{\rm FCNC}  &=& 
 - \frac{1 }{4 m_{U}^2}   
   \left[  (g_T^2+g_S^2)  \beta_{i b} \beta_{j a}^* \Big(\bar d_L^i  \gamma_\mu  d_L^j\Big)  + 2 g_T^2
(V  \beta)_{i b} (V \beta)_{j a}^*  \Big(\bar u_L^i  \gamma_\mu  u_L^j\Big)  \right] \Big(
  \bar \ell_L^a \gamma_\mu   \ell_L^b \Big)~ \no \\
  &&- \frac{1 }{
  4 m_{U}^2}   
   \left[  (2 g_T^2)  \beta_{i b} \beta_{j a}^* \Big(\bar d_L^i  \gamma_\mu  d_L^j\Big)+(g_T^2+g_S^2)  (V \beta)_{i b} (V \beta)_{j a}^* \Big(\bar u_L^i  \gamma_\mu  u_L^j\Big)   \right] \Big(
  \bar \nu_L^a \gamma_\mu   \nu_L^b \Big),
  \ea}%
where $m_U$ is the common LQ mass.
These are the only relevant operators generated at tree level
(see Ref.~\cite{Barbieri:2015yvd} for a discussion of the leading 
loop contributions in this context).

In general, the relevant low-energy observables depend on three flavorful parameters
\be
\beta_{s \tau}~,~\beta_{s \mu}~,~\beta_{b \mu}~,
\ee
and two flavor universal ones
\be
	\epsilon_{T(S)} \equiv \frac{g_{T(S)} \, m_W}{g \, m_{U}} \approx g_{T(S)} \frac{122 \, \GeV}{m_{U}}~.
\ee

\subsection{Low energy input data}

The preferred region in the model parameter space ($x_\alpha$) is determined from the $\chi^2$ function
\be
	\chi^2(x_\alpha) =  \sum_i \chi^2_i = \sum_i \frac{(\mathcal{O}_i(x_\alpha) - \mu_i)^2}{\sigma_i^2}~.
\ee
We find the best fit point which corresponds to the global minimum of $\chi^{2}$, and plot $68.3\%$~($1\sigma$), $95.5\%$~(2$\sigma$) and $99.7\%$~(3$\sigma$) C.L. regions for a given variables, after marginalizing over all other variables, requiring $\Delta\chi^{2}\equiv\chi^{2}-\chi^{2}_{\rm{min}}<\Delta_{\# \sigma}$, where $\Delta_{\# \sigma}$ are defined with the appropriate cumulative distribution function.

The experimental data used in the fit are summarized in Table~\ref{tab:FlavorFit}. They are based 
on Ref.~\cite{Greljo:2015mma}, with the following minor updates:
\begin{itemize}
\item The combination of the charged current anomalies in $B\to D^* (D)  \tau \nu$ decays (parameterized by $R_0$)~
\cite{Lees:2013uzd,Huschle:2015rga,Aaij:2015yra} includes the recent Belle measurement with semileptonic tag for the $\tau's$~\cite{belle-talk-moriond}.
\item The bounds from $B_s$ mixing  takes into account the recent Lattice QCD results on the corresponding hadronic matrix element~\cite{Bazavov:2016nty} (see also the discussion in Ref.~\cite{Blanke:2016bhf}).
\item The bound on $\Delta C_9^\mu$ is obtained from Ref.~\cite{Altmannshofer:2014rta} (and is fully consistent
with the more recent analysis of Ref.~\cite{Descotes-Genon:2015uva}).
\item The $b \to s \nu \nu$ observables have been included in the fit. In particular, we use~\cite{PDG}
\be
R_{K^{(*)}\nu}=\frac{2}{3}+\frac{1}{3}\left | 1-\frac{\Delta C^\tau_9}{C_\nu^{\rm{SM}}} \right|^2 < 4.3 \; @ ~90~\%~\rm{CL}~,
\ee 
where $C_\nu^{\rm{SM}}=-6.3$.
\end{itemize}

\begin{figure}[tbp]
  \centering
   \includegraphics[width=0.465\textwidth]{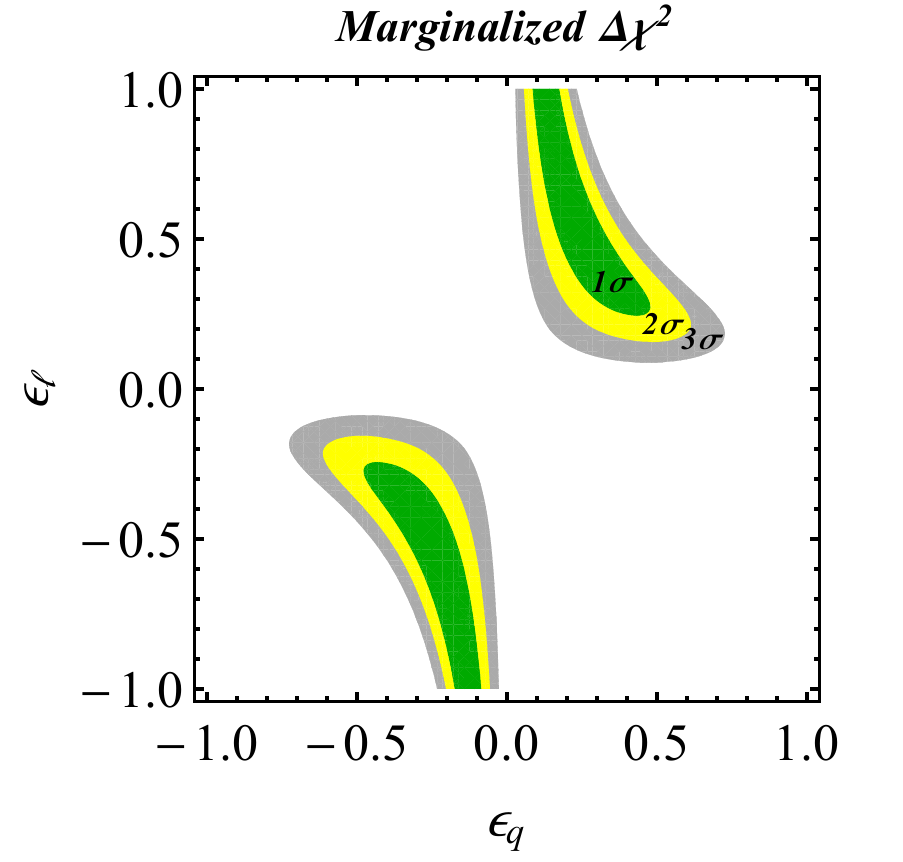} ~
   \includegraphics[width=0.5\textwidth]{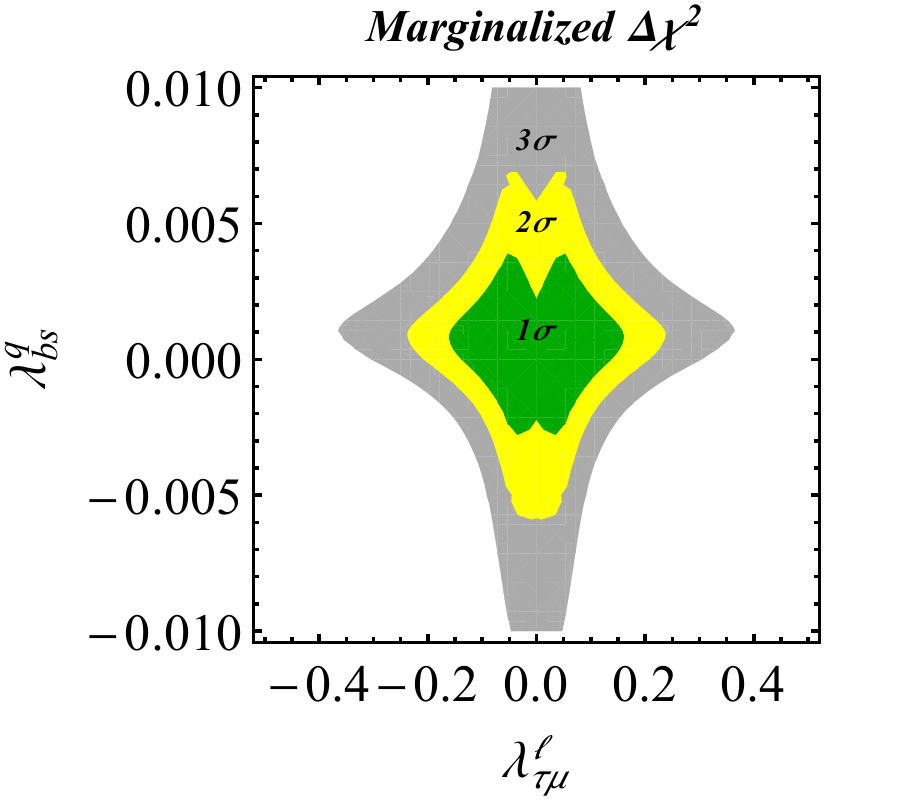}  \\
      \includegraphics[width=0.5\textwidth]{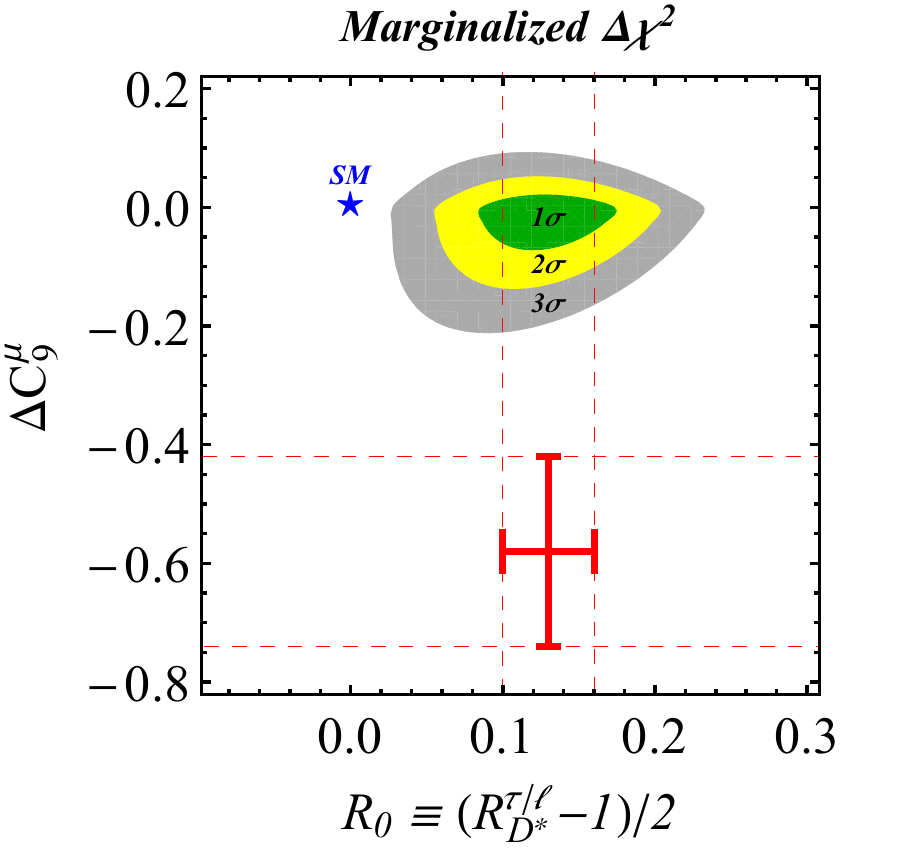}  
    \caption{ \small  {\bf Model I} ($\bar L L$ triplet + singlet only): Preferred regions at $1\sigma$ (green), $2\sigma$ (yellow) and $3 \sigma$ (gray) from the fit to low-energy data in Table~\ref{tab:FlavorFit} projected to two dimensional planes while marginalizing over other variables. Compare with $\Delta \chi^2_{\rm{SM}}=19.6$. Here $\epsilon_{q}$ ($\epsilon_{\ell}$) are the same for triplet and singlet ($\epsilon_{q,\ell}=\epsilon_{q,\ell}^{0}$).  Also, $\lambda_{\mu \mu}^\ell=| \lambda_{\tau \mu}^\ell |^2$ and $|\epsilon_{q (\ell)} | < 1$. Shown in the bottom plot in red are the experimental values for $R_0$ and $\Delta C_9^\mu$ entering the fit (see Table~\ref{tab:FlavorFit}), while the SM point is shown with the blue star.
    \label{Fig:flavor_bounds} }
\end{figure}

\subsection{Flavor fit}

Since the main features of the flavor fit are captured already in the minimal model (I), we present 
detailed numerical results only for such framework. The most significant qualitative 
differences arising in model II are briefly outlined at the end of this section.  

As discussed in Section~\ref{sec:minimal} and \ref{sec:mixing}, the vector resonances coupled to SM fermions 
in the minimal model are the $\bar L L$ states ($\rho$ and $\omega$) and the $\bar Q Q$ states 
(color-octet and color singlet, both singlets under $\SU(2)_L$). Employing the predictive flavor-mixing mechanism discussed in 
Sect.~\ref{sec:Flavor-mixing} (TC baryon mixing), the $\bar L L$ flavored vector resonances couple 
to both quark and lepton currents, while the $\bar Q Q$ resonances couple only to quarks. 
The main features arising by their contributions to low-energy observables can be summarized as follows:
 
\paragraph{($\bar L L$) states:} The results of the fit considering only $\rho$ and $\omega$ exchange are 
shown in Fig.~\ref{Fig:flavor_bounds}. In addition to the flavor relations in Eq.~(\ref{eq:minflav}), 
to further reduce the number of independent free parameters we set 
$\epsilon_{q,\ell}=\epsilon_{q,\ell}^{0}$, as expected by a  na\"ive evaluation of the relevant connected TC amplitudes.
The four parameters entering the low-energy fit are therefore 
$\{\epsilon_{q}, \epsilon_{\ell}, \lambda^q_{b s}, \lambda_{\tau \mu}^\ell \}$.
In Fig.~\ref{Fig:flavor_bounds} we show the two-dimensional planes obtained marginalizing over the remaining  variables, and requiring $|\epsilon_{q(\ell)}|<1$. Shown in green, yellow and grey are the preferred regions at $68.3\%$~($1\sigma$), $95.5\%$~(2$\sigma$) and $99.7\%$~(3$\sigma$) C.L., respectively. For comparison, the SM point has $\Delta \chi^2_{\rm{SM}}=19.6$, with respect to the best fit point.
The main conclusions are: (i) the preferred region in the 
$(\epsilon_{q},\epsilon_{\ell})$ plane is driven by the $B\to D^* (D)  \tau \nu$ anomaly ($R_0=\epsilon_{q} \epsilon_{\ell}$), as shown in the top-left plot, (ii) $\lambda_{b s}^q$ is somewhat smaller with respect to its natural expectation from the $\U(2)$ flavor symmetry ($|\lambda_{b s}^q| \sim |V_{ts}|$)
due to the $B_s$ mixing constraints (top-right plot), (iii) the discrepancy in $\Delta C_9^\mu$ can not be fully explained, mainly due to the combined constraints from 
$B_s$ mixing and LFU in $\tau \rightarrow \mu / e$  (bottom plot).
These conclusions, which are very similar to those derived in Ref.~\cite{Greljo:2015mma}, 
do not change significantly relaxing the assumption $\epsilon_{q,\ell}=\epsilon_{q,\ell}^{0}$.

\begin{figure}[tbp]
  \centering
     \includegraphics[width=0.465\textwidth]{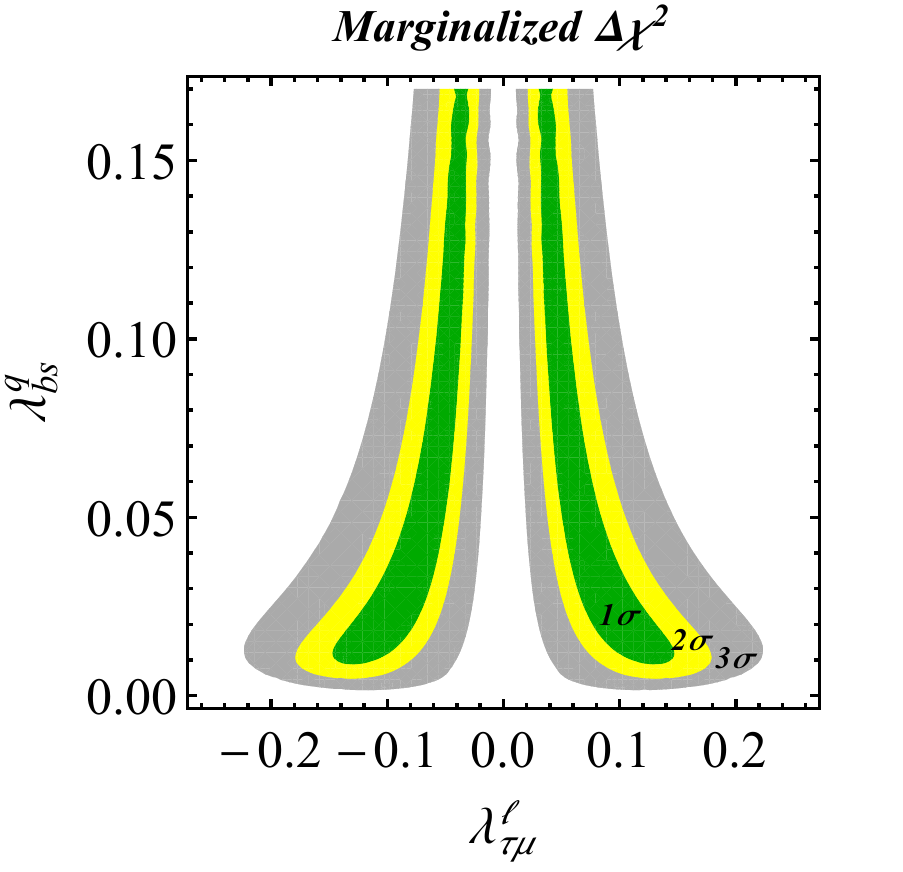}
   \includegraphics[width=0.455\textwidth]{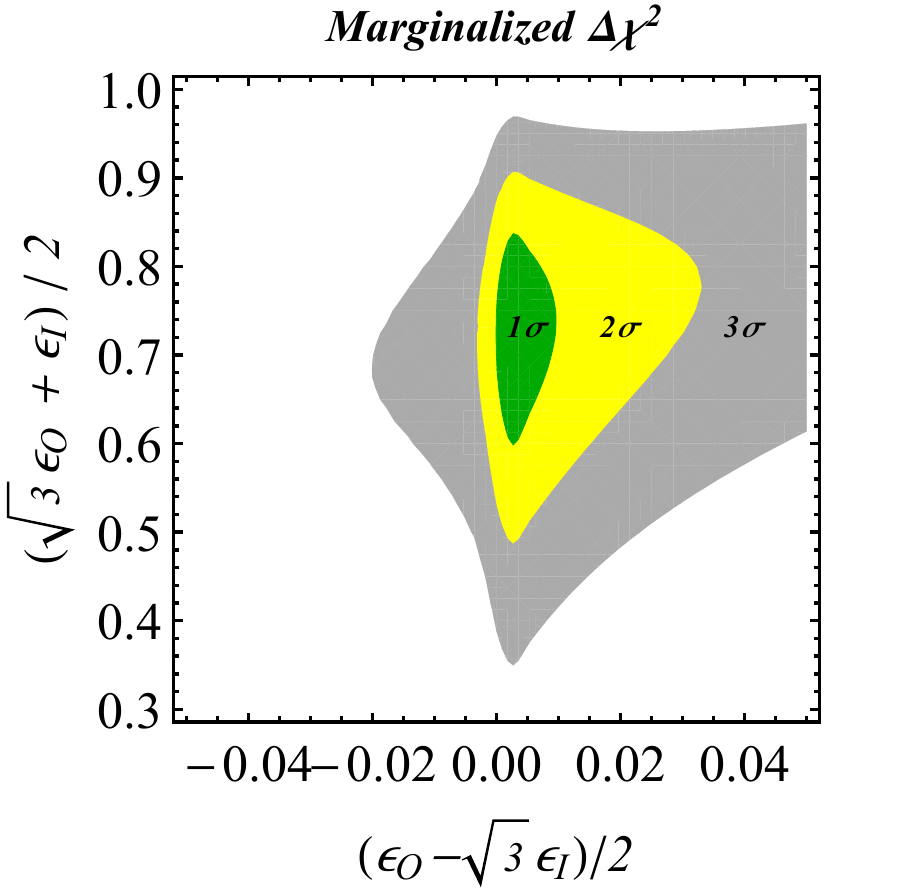}  \\ [0.5 cm]
   \includegraphics[width=0.495\textwidth]{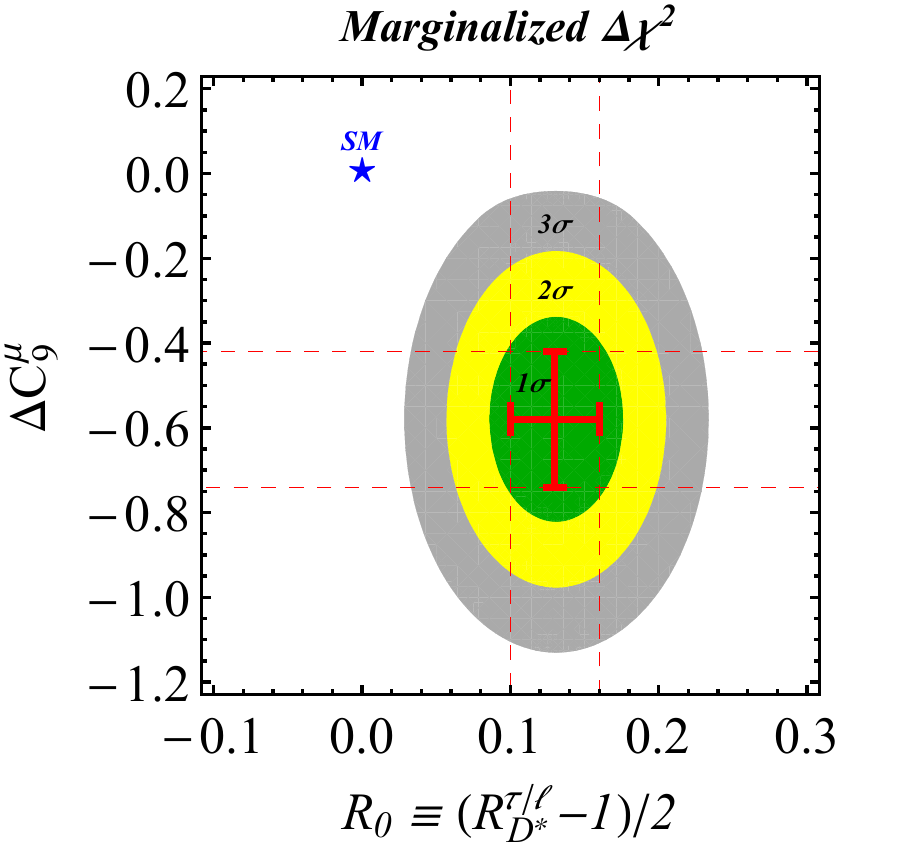} 
    \caption{ \small  {\bf Model I} ($\bar L L$ triplet + singlet and $\bar Q Q$ octet): Preferred regions at $1\sigma$ (green), $2\sigma$ (yellow) and $3 \sigma$ (gray) from the fit to low-energy data in Table~\ref{tab:FlavorFit} projected to two dimensional planes while marginalizing over other variables. Compare with $\Delta \chi^2_{\rm{SM}}=35$. Here $\epsilon_I=\epsilon_q^{(0)}=\epsilon_\ell^{(0)}$ both for triplet and singlet. Also, $\lambda_{\mu \mu}^\ell=| \lambda_{\tau \mu}^\ell |^2$ and $|\epsilon_{O(I)} | < 1$. In the bottom plot, the measurement of $R_0$ and $\Delta C_9^\mu$ from Table~\ref{tab:FlavorFit} is shown in red, while the SM point is shown with the blue star.
    \label{Fig:flavor_bounds_oct} }
\end{figure}

\paragraph{($\bar Q Q$) states:} These resonances affect only the $\Delta F=2$ amplitudes. A na\"ive evaluation of the relevant connected amplitudes points to the cancellation of the color singlet and octet contributions (see Sect.~\ref{sec:color-octet-flav}). However,  this cancellation is not protected by any symmetry and is likely to be 
violated in a realistic evaluation of the corresponding (non-perturbative) couplings in the underlying TC theory. 
To illustrate the potential impact of these resonances we perform a second fit assuming a non-vanishing coupling for the color-octet resonance ($\epsilon_{O}$). The latter should be interpreted as the net contribution resulting from the two 
sets of $Q\bar Q$ states, $\phi$ and $\mathcal{V}$.  For simplicity, we assume $\epsilon_I\equiv\epsilon_{q}=\epsilon_{\ell}$. The results are shown in Fig.~\ref{Fig:flavor_bounds_oct}.  As can be seen, with a proper tuning of the effective coupling 
$\epsilon_{O}$  it is possible to obtain a very good fit to all data. 
For comparison, the SM point has $\Delta \chi^2_{\rm{SM}}=35$, with respect to the best fit point.
Moreover, in addition to having an excellent fit of  both $R_0$ and $\Delta C_9^\mu$,
the value of  $\lambda_{b s}^q$ can be close to its natural value (top-left plot). 
However, the price to pay for this nice consistency with data 
is a non-neglible fine-tuning among two apparently unrelated  
non-perturbative TC parameters, namely  $\epsilon_{O}$ and $\epsilon_{I}$ (top-right plot).

\bigskip\bigskip

In  addition to the vector resonances discussed above, the extended model contains 
additional $\bar Q Q$ states and two vector LQ representations: $({\bf 3},{\bf 1},\Delta Y)$ and  $({\bf 3},{\bf 3},\Delta Y)$ (see Sect.~\ref{sec:leptoquark}). In model II\,A, the $\bar L L$ states behave exactly in the same way as in the minimal model, thus the previous discussion trivially applies; the $\bar Q Q$ flavored vectors contribute only to the meson mixing amplitudes, and therefore their net effect is qualitatively similar to the one occurring in the minimal case. 
In model II\,B, the role of  $\bar L L$  and $\bar Q Q$  states is exchanged, but again the qualitative picture does not 
change.

The only significant difference between the minimal and non-minimal cases are the extra contributions due to the 
LQ  states.  As shown in Ref.~\cite{Barbieri:2015yvd} by means of a pure bottom-up approach, 
these states alone can provide a viable solution to the 
$B$-physics anomalies. However, a na\"ive evaluation of the relevant TC couplings suggests that LQ triplet and singlet contributions to charged currents tend to cancel (see Eq.~\eqref{eq:lq-charged}). As  in the case of 
color-octet and color-singlet contributions to $\Delta F=2$ amplitudes, 
this cancellation is likely to be spoiled in concrete models. 
In the extended framework one can therefore conceive the possibility that 
both LQ and colorless triplet  take part in the solution of the $B\to D^* (D)  \tau \nu$ anomaly. 
The corresponding low-energy  fit has large degeneracies in the preferred parameter space,
that could be resolved only with the help of high-$p_T$ data  (see Sect.~\ref{sec:LHCpheno}). 
However, it remains true that in order to solve the discrepancy in $\Delta C_9^\mu$ and, at the same time, 
be consistent with the constraints from $B_s$ mixing and LFU in $\tau \rightarrow \mu / e$,
a $O(10\%)$ tuning among non-perturbative TC parameters is needed.

As a final comment, let us add that if more resonances contribute to the flavor observables, as generically expected in a complete model, the conclusions outlined here remain valid, but smaller couplings are allowed.

\section{LHC Phenomenology}
\label{sec:LHCpheno}

In this section we study the core phenomenology of the resonances arising in our models at the LHC. We first focus on the vector meson resonances responsible for the flavor anomalies: the colorless weak triplet and singlet, and the leptoquarks. Then, we briefly describe the main phenomenological aspects of the pNGBs, other than the singlet $\eta$.

\subsection{Vector resonances}

Vector resonances involved in solutions of $B$ decay anomalies are expected to show up in direct searches at LHC presumably in $1.5\div 2.5 \TeV$ mass window. In fact, their mass scale, $m_\rho \sim g_\rho f$, is fixed by the diphoton signal, since this fixes the pNGB decay constant $f \sim 200 \GeV$ for $N_{TC} = 3$. Here, we discuss the bounds from the existing Run-I searches, as well as the future search strategy for Run-II.

In the whole discussion we shall specify to the simple models presented in the previous sections. Notice, however, that the predictions for the cross-sections and widths of the resonances could easily change by a significant amount in a realistic scenario. This can be due to various effects: (a) the presence of further resonances -- in any case present as excited states of the vector mesons, but possibly also arising from an extended symmetry breaking pattern -- will in general enhance the meson contribution to flavor observables, thus allowing for smaller couplings to fermions; (b) the form factors associated with the TC meson-baryon couplings are expected to decrease at high energies.
The common consequence of all these effects is to reduce both the predicted cross-section for the resonances, thus relaxing the present and future LHC limits, as well as the width of the meson states, which turn out to be quite large in our minimal setup.

\begin{figure}[tbp]
  \centering
   \includegraphics[width=0.445\textwidth]{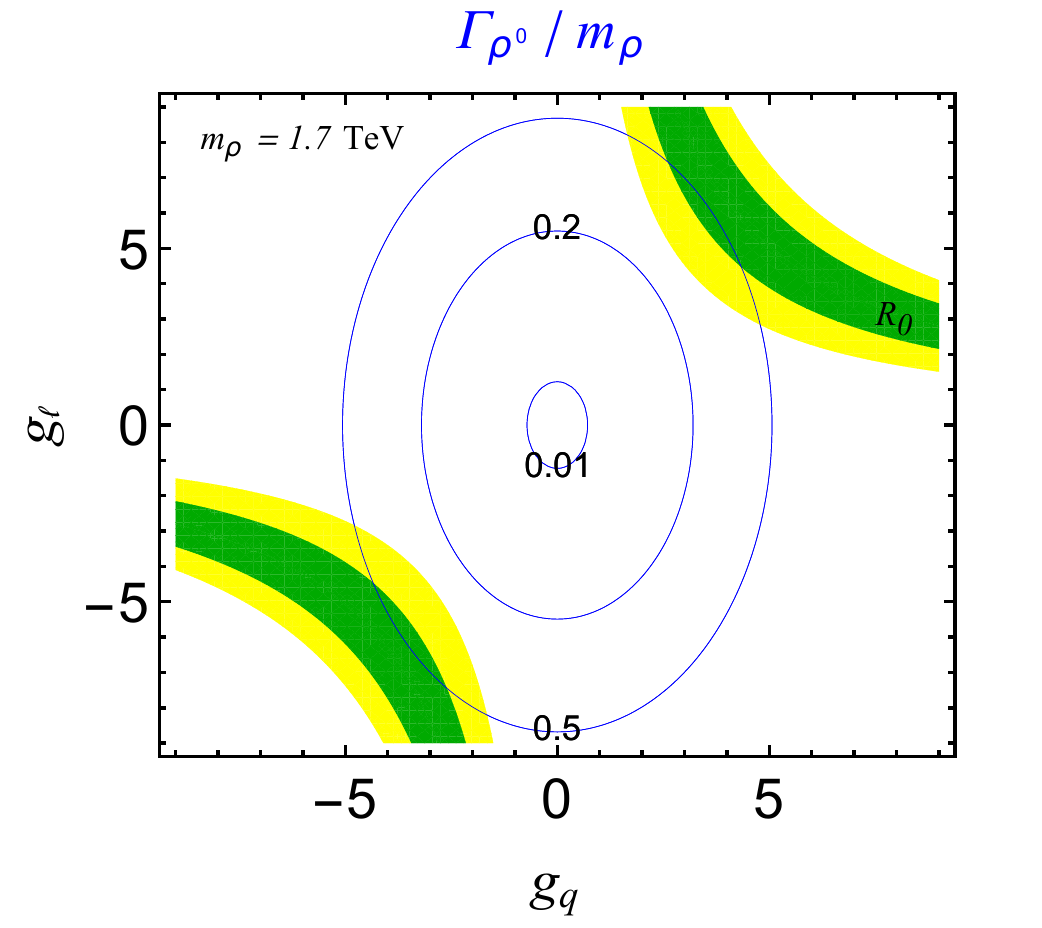} \;\;
      \includegraphics[width=0.465\textwidth]{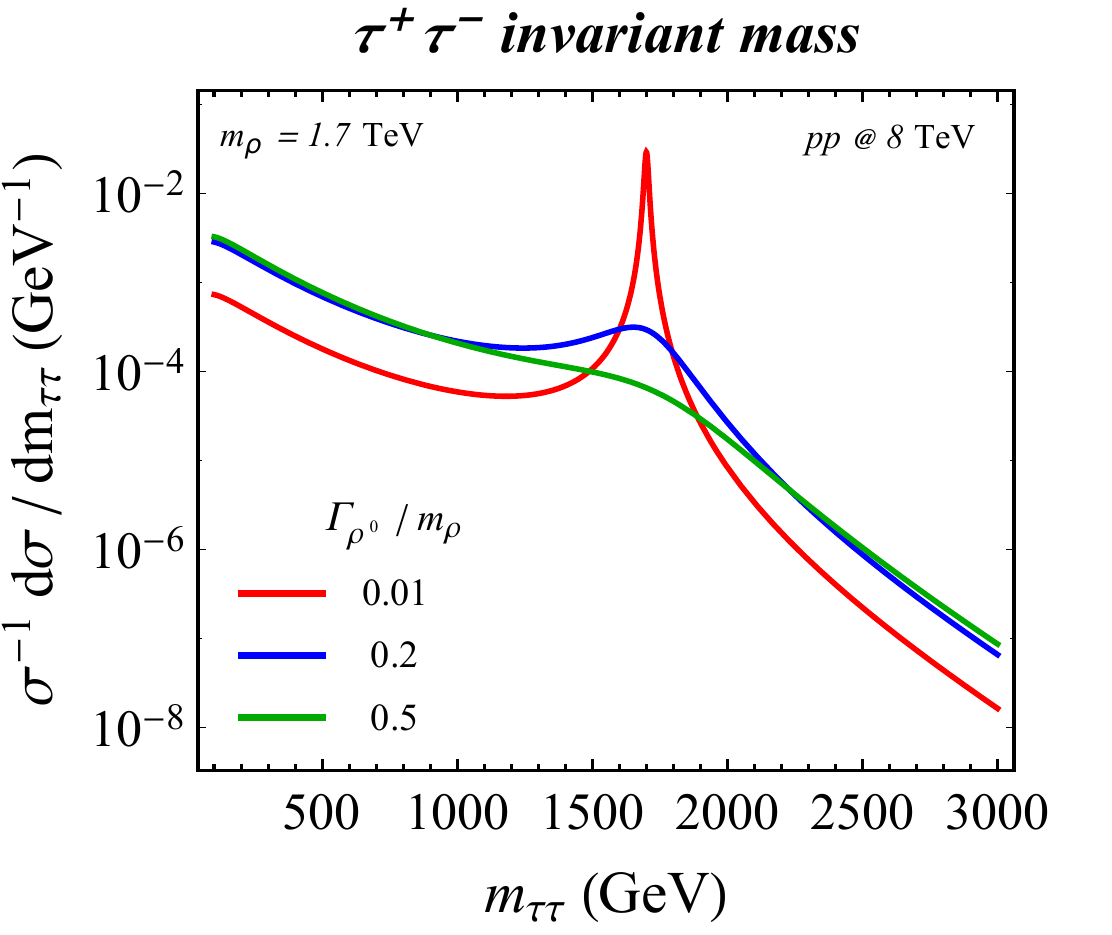} 
    \includegraphics[width=0.465\textwidth]{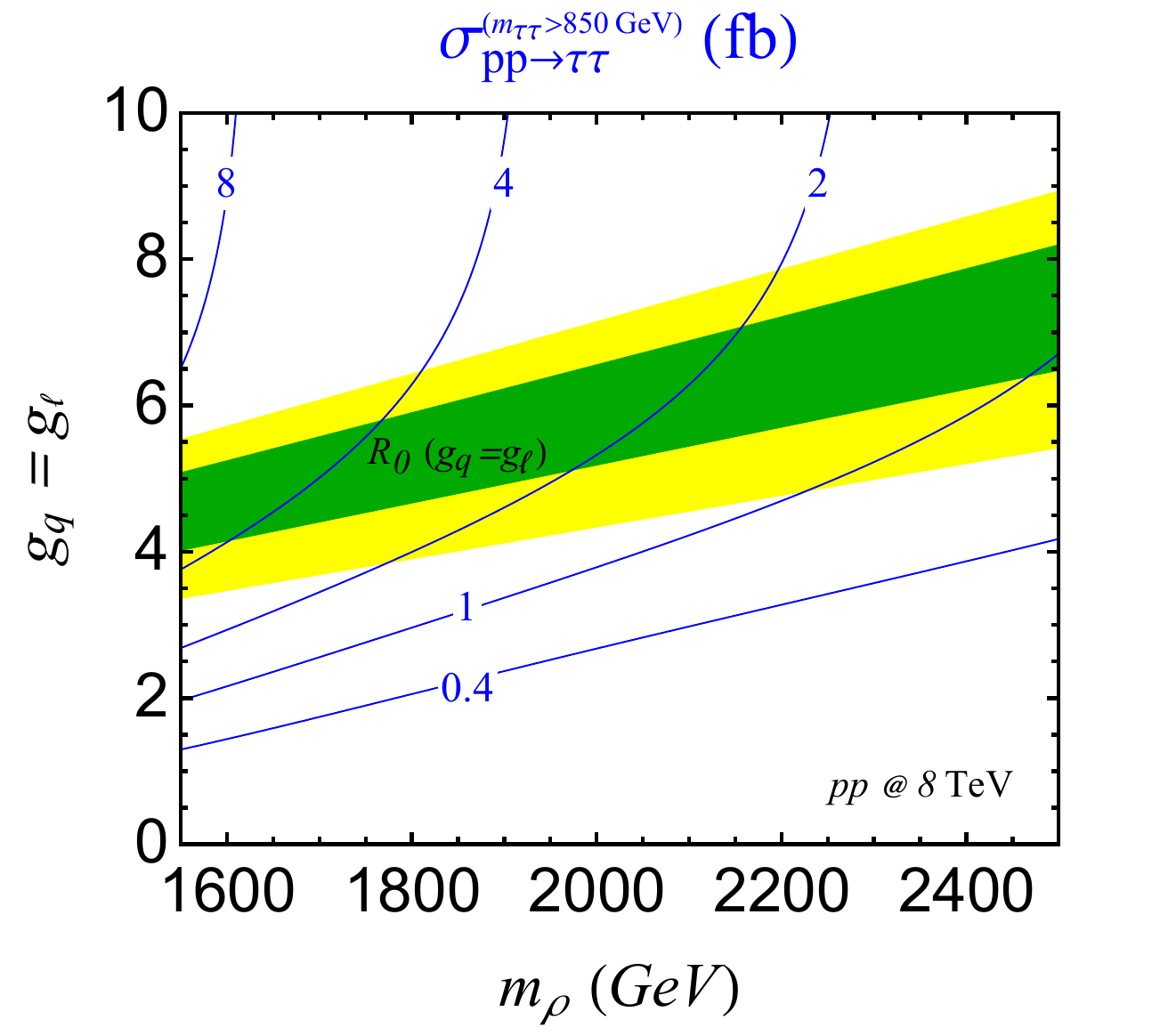} 
    \caption{ \small \label{Fig:rho_width} Top left: Total width to mass ratio isolines for $\rho^0$ (solid-blue). Shown in green and yellow is the preferred region from the $b\to c \tau \nu$ anomaly at $68\%$ and $95\%$ CL respectively, assuming the contribution from a single vector triplet resonance. Top right: Normalized $\tau\tau$ invariant mass distribution in $p p \to \tau^+ \tau^-$ at 8 TeV for $m_\rho=1.7$~TeV and width to mass ratio; 0.01~(red), 0.2~(blue) and 0.5 (green). Bottom: $p p \to \tau^+ \tau^-$ production cross section at 8~TeV for $m_{\tau \tau} > 850$~GeV assuming $\rho^0$ contribution only and  $g_q=g_\ell$ (solid blue). The preferred region from $b\to c \tau \nu$ assuming $g_q=g_\ell$ is shown in green and yellow.}
\end{figure}

\paragraph{$\SU(2)_L$ triplet $\rho_\mu^a$:}
The vector triplet $\rho_\mu^a\equiv({\bf 1},{\bf 3},0)$ is in direct connection with the $b \to c \tau \nu$ anomaly, and its LHC phenomenology has been discussed in Ref.~\cite{Greljo:2015mma}. Here, we extend the discussion to the relevant high-mass region. The resonances ($\rho^0$ and $\rho^{\pm}$) primarily decay to third generation fermions. In particular, the partial widths of the neutral component ($\rho^0$) are
\be
\Gamma(\rho^0 \to \tau^+ \tau^- ~(\nu_\tau \bar \nu_\tau)) = \frac{g_\ell^2}{96 \pi} m_{\rho}\,,\qquad\qquad \Gamma(\rho^0 \to b \bar b~ ( t \bar t)) = \frac{g_q^2}{32 \pi} m_{\rho}\,.
\ee
The decay to a pair of triplet pNGB, $\pi^a \sim \bar{L} \sigma^a L$, is expected to be the dominant channel in such strongly-coupled models. However, due to the large couplings to third generation fermions required by the flavor fit, it is sub-leading (or at most of the same order) as the ones discussed above. Parametrizing the interaction as $\cL^{\rho\pi\pi} = \frac{1}{2} g_{\rho\pi\pi} \epsilon^{abc} \rho_\mu^a \, \pi^b \partial_\mu \pi^c$, with $g_{\rho\pi\pi} \lesssim 4\pi$, one finds
\be
	\Gamma(\rho \to \pi\pi) = \frac{g_{\rho\pi\pi}^2}{192 \pi} m_\rho \left( 1 - \frac{4 m_\pi^2}{m_\rho^2} \right)\,.
\ee
Finally, the decay width into SM vector bosons is small if the Higgs field is not directly coupled to the TC sector, as is the case in our setup, and only arises at the one-loop level.
In Fig.~\ref{Fig:rho_width} (top left) we show the total width of $\rho^0$ into fermions (blue lines) in the $(g_q,g_\ell)$ plane, normalized to its mass fixed at $m_\rho = 1.7\TeV$.
We conclude that, in the region relevant for the $R_0$ anomaly, the resonance is very wide with $\Gamma/m_\rho \sim 30 \div 60 \%$. This can however easily be reduced by a factor of a few in a more complete model.

As shown in Ref.~\cite{Greljo:2015mma}, the main production mechanism at a hadron collider is $b \bar b \to \rho^0$.
In the standard vectorlike confinement setup, the $\rho^a_\mu$ resonances mix with the SM gauge bosons $W^a_\mu$ with a mixing angle $\sim g / g_\rho$, inducing diagonal couplings to light SM quarks which determine the production cross section at the LHC~\cite{Kilic:2010et} via Drell-Yan. In our case, instead, the bottom fusion production mechanism is the dominant one. The ratio of $b \bar b$- to $u \bar u$-induced cross sections is
\be
R_{b\bar b / u \bar u } \approx \frac{g_q^2}{(g^2/g_\rho)^2}~\frac{\mathcal{L}_{b \bar b}}{\mathcal{L}_{u \bar u}}~,
\ee
where $\mathcal{L}_{b \bar b,u \bar u}$ are the respective parton luminosities. For example, fixing $g_q \sim 5$ and $m_\rho \sim 1.7$~TeV, consistently with the flavor fit, we find $R_{b\bar b / u \bar u } \sim 7$. In the numerical results presented in this section, we convolute the analytical formulas derived in Appendix~\ref{App:2} with the MMHT 2014 NNLO parton distribution functions (PDF)~\cite{Harland-Lang:2014zoa}. All the results have also been checked with MadGraph~\cite{Alwall:2014hca} implementing the model in FeynRules~\cite{Alloul:2013bka}.

The best present limits on the model are due to the ATLAS search for heavy $Z'$ decaying to $\tau^+ \tau^-$~\cite{Aad:2015osa}, which sets the limit $\sigma(p p \to Z' + X) \times \mathcal B(Z' \to \tau^+ \tau^-) \lesssim 4 ~(7)$~fb assuming a narrow (moderate) width resonance in the mass window 1.5---2.0~TeV. 
In Fig.~\ref{Fig:rho_width} (top right), we show the normalized differential cross section in the $\tau \bar{\tau}$ invariant mass in $p p \to \tau^+ \tau^-$ at 8 TeV, for three cases: (a) narrow width ($\Gamma_{\rho^0}/m_\rho=0.01$), (b) moderate width ($\Gamma_{\rho^0}/m_\rho=0.2$), and (c) wide resonance ($\Gamma_{\rho^0}/m_\rho=0.5$), while fixing $m_\rho=1.7$~TeV. Obviously, the limits on a narrow (moderate) width resonance cannot trivially apply for (c), which is the relevant case for the flavor anomalies.
Moreover, the final event selection in Ref.~\cite{Aad:2015osa} is based on the total transverse mass variable ($m_{\rm{T}}^{\rm{tot}}$), which is optimized for a narrow resonance search. In particular, for $m_{Z'} > 1.5$~TeV, $m_{\rm{T}}^{\rm{tot}}$ is required to be grater than $850$~GeV in the signal region (See Table~4 in~\cite{Aad:2015osa}). An appropriate recast of the analysis requires good control over $\tau$ reconstruction, which is beyond the scope of this paper. In order to qualitatively estimate the present bound on our $\rho^0$, we crudely approximate $m_{\rm{T}}^{\rm{tot}} \sim m_{\tau\tau}$ and plot in Fig.~\ref{Fig:rho_width} (bottom) the total $p p \to \tau^+ \tau^-$ cross section at $8$~TeV 
requiring $m_{\tau \tau}^{\rm{min}}>850$~GeV, and setting $g_\ell = g_q$. Interestingly enough, the present searches are just starting to probe the relevant parameter region as predicted by the diphoton and flavor anomalies. 
Clearly, future search strategies for the $\rho^0$ in this model should be optimized for a wide resonance.

Another important decay channel is the one into a pair of $b$ quarks. Since the flavor fit constrains the product $\epsilon_\ell\,\epsilon_q$, this channel can turn out to be relevant if the quark coupling is larger than the lepton coupling. The CMS search for narrow resonance $G$ decaying into ($b$-enriched) dijets \cite{Khachatryan:2015sja} at present constrains the cross-section $\sigma(pp\to G + X)\times \mathcal{B}(G\to jj)\times \mathcal{A}\lesssim 150\,{\rm fb}$ (where $A$ is the acceptance) at a resonance mass of 1.9 TeV, where it shows an excess at the level of about $2.5\sigma$. These limits are significantly weakened in the case of a broad resonance.

A few comments are in order for the specific model realization discussed above. In model I (Sect.~\ref{sec:minimal}), with the flavor structure generated via mixing with TC baryons (Sect.~\ref{sec:mixing}), the contribution from the vector singlet resonance $\omega_\mu\equiv ({\bf 1},{\bf 1},0)$ is expected to interfere constructively with the $\rho^0$ one, enhancing the cross section by a factor $\sim 4$ with respect to Fig.~\ref{Fig:rho_width} (bottom). If, instead, the baryon-mixing assumption is relaxed (Sect.~\ref{sec:non-mixing}), the interference pattern is arbitrary and a destructive one would also be possible.

To summarize, a positive signal in $\tau^+ \tau^-$ (or $b\bar b$) searches is a generic prediction if a vector triplet resonance ($\rho^a_\mu$) is involved in the solution of $R_0$ anomaly.

\paragraph{Vector leptoquarks $\mathcal{U}^a_\mu$ and $U_\mu$:}
Let us now consider the case in which a vector leptoquark ($\mathcal{U}^a_\mu\equiv({\bf 3},{\bf 3},2/3)$, or $U_\mu\equiv({\bf 3},{\bf 1},2/3)$) is responsible for the $b \to c \tau \nu$ anomaly, a possibility discussed in the context of the second model. The large coupling to bottom quark and tau lepton, required to fit the excess, also gives rise to sizable $b \bar b \to \tau^+ \tau^-$ signal at the LHC via $t-$channel LQ exchange (see Appendix~\ref{App:2}).

In the limit $m_{U} \gg  m_{\tau \tau}/2$, the tau leptons angular distribution is similar to the one in the $s-$channel production. Otherwise, in the opposite limit, tau leptons are very forward and can easily escape the detection. Interestingly, in our case the hadronic cross section is usually saturated at low $m_{\tau \tau}$, implying that the contribution to the signal region is expected to be as sizable as before. For example, fixing $m_{U}=1.7$~TeV and requiring $R_0=0.13$ (which translates to $g_{T(S)}=7.1$) one finds $\sigma(p p \to \tau^+ \tau^-) \sim 10$~fb for $m_{\tau\tau}\ge850$~GeV. Naively comparing with the quoted excluded cross sections for the narrow s-channel resonance~\cite{Aad:2015osa}, a vector leptoquark that fully saturates $R(D^*)$ is also expected to show up very soon in LHC $\tau^+ \tau^-$ searches.

It is interesting to point out that in the case of comparable $U_\mu$ and $\rho^0_\mu$ contributions to the $b\to c \tau \nu$ anomaly, with fine adjustment of the parameters one can achieve destructive interference between $s-$ and $t-$ channel diagrams shown in Fig.~\ref{Fig:diagrams}, lowering the total cross section even by an order of magnitude.

\subsection{Pseudoscalars}

\paragraph{$\SU(2)_L$ triplet $\pi^a$:}
For $m_L < m_Q$ the lightest pNGB multiplet in both models is a $\SU(2)_L$ triplet $\pi^a \sim ({\bf 1}, {\bf 3}, 0)$, with mass $m^2_\pi \simeq 2 B_0 m_L + \Delta m^2_{({\bf 1},{\bf 3},0)}$, where the gauge contribution is shown in Eq.~\eqref{eq:gauge_mass_contr}. EWSB effects are expected to further generate a small mass splitting inside the triplet, increasing the mass of the charged ones, as for the QCD pions. The lightest techni-hadron in these models, therefore, is likely to be the neutral component $\pi^0$.
As for the singlet $\eta$, also this triplet is expected to couple to the SM fermion doublets of the third generation via strong interactions with two baryons, Eq.~\eqref{eq:eta_bar_Lagr}:
\be
	\Delta \cL = i g_{\pi B B} \pi^a ( \bar{B}_\ell \gamma^5 \sigma^a B_\ell + \bar{B}_q \gamma^5 \sigma^a B_q).
\ee  
After baryon-SM fermion mixing, Eq.~\eqref{eq:Btoq}, the most relevant couplings generated are
\be
	\Delta \cL = i \frac{g_{\pi B B} \kappa_q^2 m_t}{m_B} \left( \pi^0 \bar{t} \gamma^5 t + \frac{1}{\sqrt{2}} \pi^+ \bar{t}_L \gamma^5 b_R + \frac{1}{\sqrt{2}} \pi^- \bar{b} \gamma^5 t \right)~.
	\label{eq:}
\ee
Contrary to the $\eta$ coupling to $B_q$ in model I, the pions do not have a suppression of $\sqrt{15}$, thus $g_{\pi B B}\sim g_\rho$. For this reason, and given the absence of an anomalous coupling to two gluons, the decay to $t \bar{t}$ is expected to be the dominant one for the neutral pion $\pi^0$. A subleading decay mode is due to the anomalous coupling to two photons (as well as to the other EW gauge bosons), given by $A_{\gamma\gamma}^{\pi^0} = 2 N_{TC} Y_L$, which induces a decay width
\be
	\Gamma(\pi^0 \to \gamma\gamma) = \frac{N_{TC}^2 \alpha^2 Y_L^2}{16 \pi^3} \frac{m_{\pi^0}^3}{f^2}.
\ee
The main production mechanism for these states at the LHC is pair production via Drell-Yan, $q \bar{q} \to W^{\pm *} \to \pi^{\pm} \pi^0$ or $q \bar{q} \to Z^{*}/\gamma^* \to \pi^{+} \pi^{-}$. The electroweak production mode, the mass $m_\pi \gtrsim 400 \GeV$, see Fig.~\ref{Fig:spectrum}, and the decay channels with many tops in the final state make the search for these states quite challenging at the LHC.

\paragraph{Color octets $\tilde \pi$:}

\begin{figure}[t]
  \centering
     \includegraphics[width=0.5\textwidth]{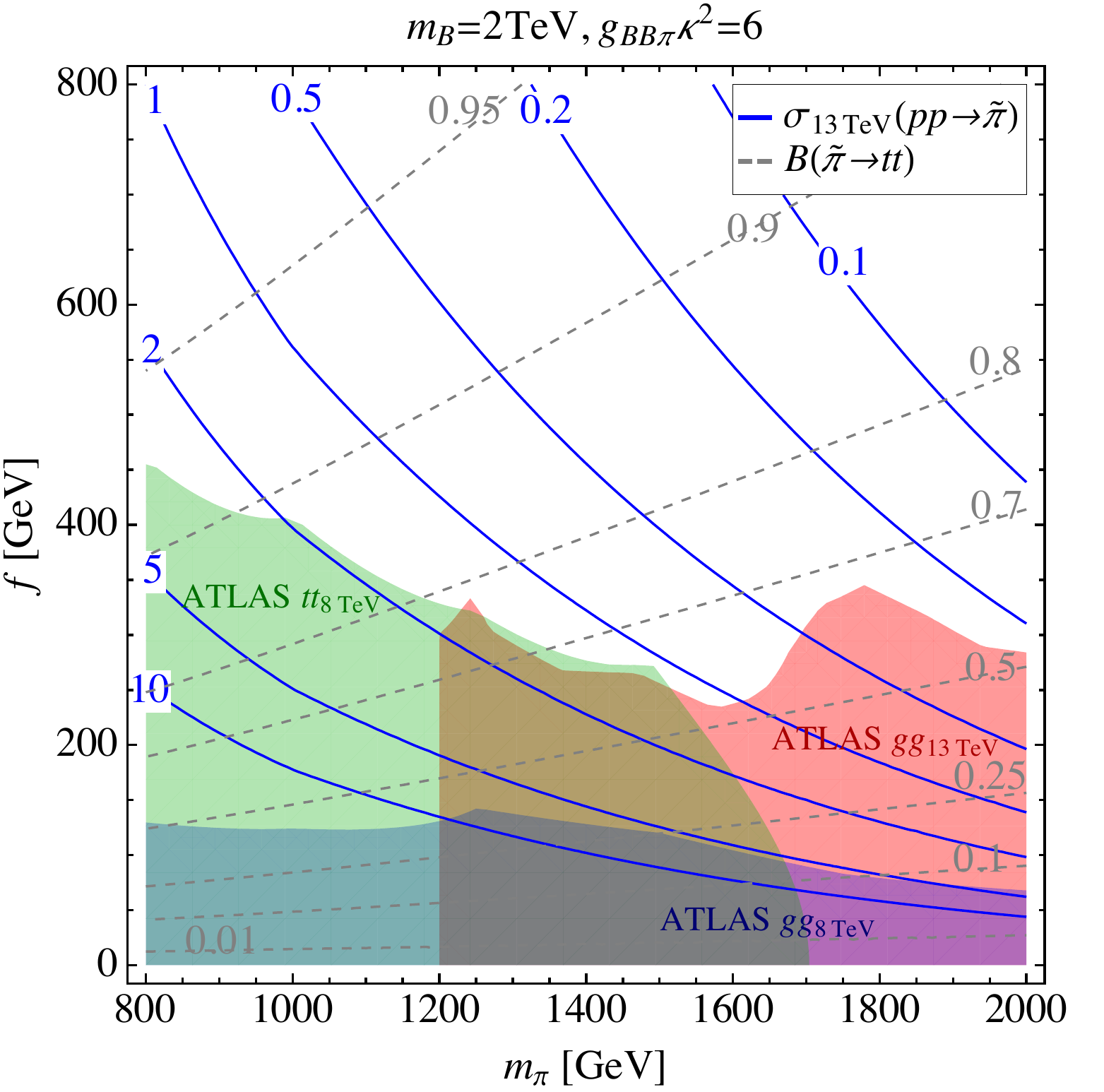}
    \caption{ \small  Experimental bounds on the color-octed $\SU(2)_L$-singlet pseudoscalar $\tilde{\pi}$ from resonant searches in $t\bar{t}$ (8TeV) \cite{Aad:2015fna} and in dijet (both at 8TeV \cite{Aad:2014aqa} and 13TeV \cite{ATLAS:2015nsi}), in the plane of the pseudoscalar mass vs. the decay constant $f$. The color-octet coupling to $t\bar t$ is obtained by fixing the baryon mass and the coupling, analogous to those in Eq.~\eqref{eq:eta_ttbar_Lagr},  to $m_B = 2.0$ TeV and $g_{\tilde{\pi}BB} \kappa_q^2 = 6$. The blue lines are isolines of the $\sigma(pp \to \tilde{\pi})$ production cross section at 13TeV (in pb), while the dashed gray lines are isolines of the branching ratio of the decay in $t\bar{t}$.}
    \label{Fig:LHC_bounds_oct}
\end{figure}

A discussion of the phenomenology of the color-octets pNGB, $\tilde \pi_1 \sim ({\bf 8}, {\bf 1}, 0)$ and $\tilde \pi_3 \sim ({\bf 8}, {\bf 3}, 0)$, can be found for example in Refs.~\cite{Craig:2015lra,Redi:2016kip,Bai:2016czm}. Their mass is given by $m^2_{\tilde \pi_{1(3)}} = 2 B_0 m_Q + \Delta m^2_{({\bf 8},{\bf 1(3)},0)}$, thus for $m_Q > m_L$ they are expected to be the heaviest pNGB with $m_{\tilde \pi} \sim 1.2 \div 1.5 \TeV$, see also Fig.~\ref{Fig:spectrum}.
It is worth stressing that the presence of such a state is a model-independent prediction of all models trying to explain the diphoton excess in terms of a singlet pNGB coupled to gluons via the anomalous coupling. Since this implies the presence of some fundamental TC fermion $Q$ charged under $\SU(3)_c$, then a color-octet pNGB, $\tilde \pi^A \sim (\overline Q T^A Q)$, will always be present in the spectrum \cite{Bai:2016czm}.

While both the $\SU(2)_L$ triplet and singlet can be doubly produced in gluon fusion, the singlet also can be singly produced via its anomalous couplings to two gluons. This coupling does not depend on the details of the specific models: it is only a function of $N_{TC}$ and the pion decay constant $f$, see e.g. \cite{Redi:2016kip,Bai:2016czm}. In the analysis of Ref.~\cite{Bai:2016czm} it was shown how the present bounds from dijet searches, both at 8 TeV \cite{Aad:2014aqa,Khachatryan:2015sja} and at 13 TeV \cite{ATLAS:2015nsi} already cast strong bounds on such a state. It was also shown that the anomalous couplings to $g \gamma$ are phenomenologically relevant and, for some models, can give comparable bounds as those from dijet searches.
The main difference between our setup and those considered in previous studies is the strong coupling of this particle with a $t \bar{t}$ pair via the mixing with baryons, which we can parametrize with an analogous Lagrangian as was done for the $\eta$ in Eq.~\eqref{eq:eta_ttbar_Lagr}. This term opens up the decay channel to $t \bar{t}$, which we find to be comparable in size to the one in two gluons.
In Fig.~\ref{Fig:LHC_bounds_oct} we show the experimental bounds on such a state from the ATLAS 8 TeV resonant search in $t\bar{t}$ \cite{Aad:2015fna,CMS:lhr}, as well as from the ATLAS 8 TeV and 13 TeV dijet searches. In order to extract the excluded cross section from the two dijet analyses, we perform a MonteCarlo simulation to estimate the acceptance of the cuts applied to be $\cA \sim 54\%$ for both analyses. The $t\bar{t}$ branching ratio isolines are shown with dashed-gray lines while the solid blue lines show the production cross section of the pseudoscalar in gluon fusion in pb.\footnote{The production cross section is computed analytically as in Ref.~\cite{Redi:2016kip}, which we also multiply by a QCD NLO $k$-factor $k_{\rm NLO} \simeq 2$.} In the relevant parameter space, $f\gtrsim 200 \GeV$ and $m_{\tilde{\pi}} \sim 1.5 \TeV$, the total decay width in this region is $\Gamma_{\tilde \pi} \sim O(1) \GeV$ and the decay widths in $t \bar{t}$ and $gg$ are comparable. The combination of the $t\bar{t}$ 8TeV search and the dijet 13TeV search already puts some tension in the models considered here. We expect that a future update of both searches will further improve the reach on this class of models, in particular a $t\bar{t}$ resonant search at the LHC Run-2 will have a strong exclusion (or discovery) power of this kind of state.

\paragraph{Leptoquarks $\mathcal{D}$, $\mathcal{S}$, and $\mathcal{T}^a$:}
The minimal model I contains a pNGB multiplet with SM quantum numbers $\mathcal{D} \sim |\bar{L}Q\rangle \sim ({\bf 3},{\bf 2}, \Delta Y)$, where $\Delta Y = Y_Q - Y_L = - \frac{1}{3}$ ($\frac{1}{6}$) with the hypercharge assignment A (B) and masses $m^2_{\mathcal{D}} = B_0(m_L + m_Q) + \Delta m^2_{({\bf 3},{\bf 2},\Delta Y)} \sim (1 \TeV)^2$.
These states can be pair-produced at the LHC through gluon fusion and, in the minimal version of the model presented in this paper, cannot decay into SM particles. This can be easily understood by noticing that they carry baryon and lepton numbers $B = \frac{1}{6}$ and $L=-\frac{1}{2}$, which cannot be reproduced by any combination of SM states.
If the leptoquarks do not decay, they will hadronize and form stable uncolored states. If the lightest of these states is neutral, as is e.g. the case for $|\mathcal{D} \bar q_L\rangle$ in model I\,B, it could even be a Dark Matter candidate. If, on the contrary, stable charged states are present in the spectrum, this constitutes a problem for cosmology and calls for an extension of the model.\footnote{A decay channel for the leptoquarks can be opened e.g. allowing a tiny baryon number violating coupling to SM fermions, or introducing further light DM states.}
This, however, goes beyond the purpose of the present paper and we leave a more detailed study of this issue to future work.

In model II, the $\SU(2)_L$ singlet $\mathcal{S} \sim ({\bf 3},{\bf 1}, \frac{2}{3})$ and triplet $\mathcal{T}^a \sim ({\bf 3},{\bf 3}, \frac{2}{3})$ pNGB leptoquarks have masses $\gtrsim 1 \TeV$ and decay dominantly to third generation fermions through baryon-SM fermion mixing, as already discussed above. In particular, the relevant decays are $\mathcal{S}, \mathcal{T}_{2/3} \to t \bar{\nu}_\tau$, $\mathcal{T}_{5/3} \to t \bar{\tau}$ and $\mathcal{T}_{-1/3} \to b \bar{\nu}_\tau$.
The present limits from LHC on QCD pair-produced third generation LQs exclude $m_{\rm{LQ}}\lesssim 700 \div 750$~GeV~\cite{Khachatryan:2015bsa,Khachatryan:2014ura}, and are therefore still not sensitive to our setup.

\section{Conclusions}
\label{sec:conclusions}

The Standard Model successfully describes fundamental interactions over a wide range of energies.
So far no conclusive evidence indicating the need of new degrees of freedom around 
the TeV scale has been found; however, recent LHC data, both at high $p_T$ and in precise 
flavor-physics measurements,  show increasing tensions with the SM predictions.

In this paper we have proposed a coherent framework for the interpretation of the 
two largest set of anomalies: the 750~GeV diphoton excess and the hints of deviations from Lepton 
Flavor Universality observed in $B$ decays. These two phenomena 
could be two correlated manifestations of the lowest-lying resonances of
a new strongly-coupled sector with vectorlike confinement.

The interpretation of the 750~GeV diphoton excess as the decay, via the anomaly,  
of a pNGB of a new strongly interacting sector has already been widely discussed 
in the literature. We have shown that the same framework can provide a natural ultraviolet 
picture for the effective theories with heavy vector mediators (both color singlets and leptoquarks) 
proposed to explain the evidences of LFU violation in $B$ decays.
We have presented and analyzed two explicit examples of such models, with different particle content
and number of free parameters. Already in the minimal model, that predicts 
no leptoquark contributions to $B$ decays, it is possible 
to obtain a very good fit of both low- and high-energy data. In particular, we find that the solution of the 
anomaly in $B\to D^{(*)} \ell \nu$ decays is quite robust and does not require specific tuning 
of the parameters of the model. 

The explicit models discussed in this paper should be considered as representative examples
to illustrate the main features of this set-up and the possible connections between 
low- and high-energy observables. More detailed models, addressing 
the origin of flavor mixing and the connection with the Higgs sector, are beyond the scope of the 
present analysis.
Despite the large number of free parameters (and the arbitrariness in the choice of the new gauge interaction 
and fundamental TC constituents), we have shown that the overall framework leads to a few basic predictions.  
On general grounds, heavy vector resonances with large widths 
and dominant coupling to third-generation quarks and leptons should be expected in the 1.5--2.5~TeV mass range. 
Their search is not easy, given the smallness of the coupling to first and second generation fermions
and the large decay width, but is well within the reach of the LHC Run-II at 13 TeV. A particularly   promising channel is the inclusive $\sigma(p\,p\to\tau^+\tau^-)$, 
where a significant excess over the SM is expected in the tail of the $m_{\tau\tau}$ distribution.
Interestingly, such an excess is an unambiguous prediction following the solution of the 
$B\to D^{(*)} \ell \nu$ anomaly, independently of the type of meditator (color-singlet or leptoquark exchange). In addition, searches in the $pp\to b\bar b$ channel could also be relevant for the vector mesons.
We therefore encourage the experimental collaborations to put an effort into testing this picture during the present LHC run.

Should the 750 GeV diphoton excess be disfavored as more data is collected by the LHC experiments, the framework presented in this paper could still be considered as a viable explanation of the flavor anomalies, as due to the exchange of heavy vector TC mesons. In this case the scale of the theory would remain in the same ballpark, being fixed by the mass of these states, while the masses of the various pNGB, including the singlets, would depend on the unknown TC fermion masses. Such a scenario would still motivate searches for spin-0 resonances in all the diboson channels discussed here, including the diphoton one.

If this class of models is indeed realized in nature, the diphoton excess and the flavor anomalies would be the first glimpses of a new spectroscopy, opening up a very rich research program for both high-$p_T$ LHC searches and precision flavor-physics measurements.

\section*{Acknowledgments}

This research was supported in part by the Swiss National Science Foundation (SNF) under contract 200021-159720. We thank Roberto Contino, Jernej F. Kamenik, and Jonas Lindert for useful discussions. \\

\noindent
{\bf Note Added: }
New data presented at ICHEP in August 2016 by ATLAS and CMS \cite{ATLAS:2016diphoton_ICHEP,CMS:2016diphoton_ICHEP} do not show any sign of the diphoton excess discussed in this work.
While this reduces the interest of some of the phenomenological aspects presented here, the discussion about the flavor anomalies and their interpretation in the context of strongly coupled models remain relevant.
As outlined in the conclusions, light spin-0 resonances are a generic prediction of this class of models, providing a strong motivation for their further search at the LHC in all the relevant channels.

\appendix

\section{Updated diphoton signal-strength combination}

\begin{figure}[tbp]
  \centering
   \includegraphics[width=0.55\textwidth]{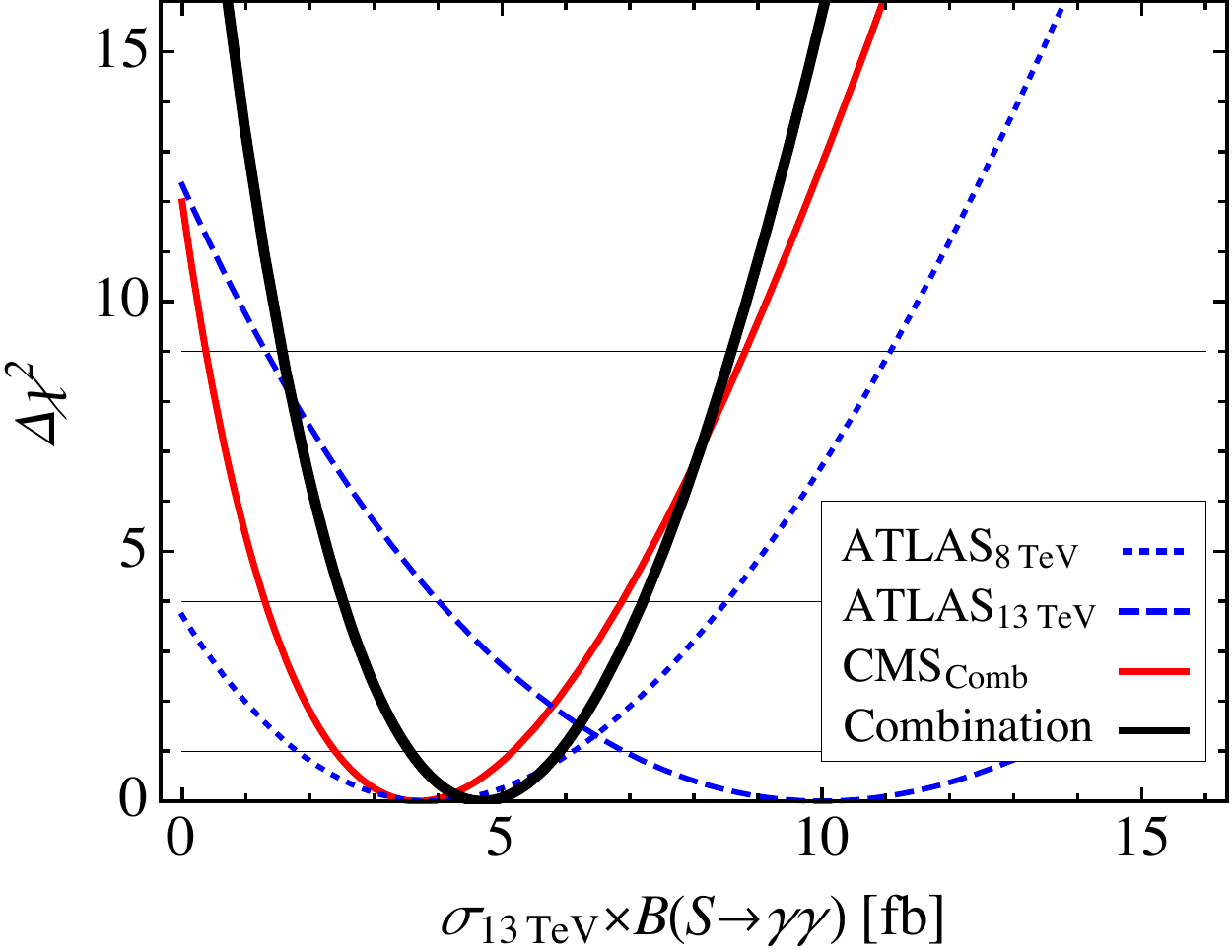}  
    \caption{ \small  Combination of the diphoton cross section from ATLAS \cite{ATLAS:2016diphoton} and CMS \cite{CMS:2016owr} with both 8 and 13 TeV data, interpreting the resonance as a narrow-width pseudoscalar produced in gluon fusion.
        \label{Fig:diphoton_comb} }
\end{figure}
Our combination of the most recent analysis of the diphoton excess performed by ATLAS \cite{ATLAS:2016diphoton} and CMS \cite{CMS:2016owr}, is shown in Fig.~\ref{Fig:diphoton_comb}. We use the published  CMS combination of the 8~TeV and 13~TeV data (red solid line), while the ATLAS 8 TeV (dotted blue) and 13~TeV (dashed blue) constraints are extracted from the available information by assuming they follow a Poisson distribution, as already done in Ref.~\cite{Buttazzo:2015txu}. The 13~TeV distribution is obtained by imposing that a 3.5$\sigma$ significance (for the narrow-width case) and an observed excluded cross section at 95\%CL of $\sigma^{\rm excl}_{\rm 95\% CL} = 10/\epsilon_{\rm acc}$~fb are reproduced, where $\epsilon_{\rm acc} \sim 0.6$ is our estimate of the acceptance in the fiducial region \cite{Buttazzo:2015txu}. The 8~TeV distribution is instead obtained imposing a 1.9$\sigma$ significance and a compatibility with the 13~TeV data of $1.2\sigma$ assuming production via gluon-fusion. The final combination is shown in Fig.~\ref{Fig:diphoton_comb} (black solid line), and corresponds approximately to
\be
	\sigma_{13\TeV}(pp \to \eta) \times \mathcal{B}(\eta \to \gamma\gamma) = 4.7^{+1.2}_{-1.1}~\text{fb} ~.
	\label{eq:diphoton_comb}
\ee

\section{ $b ~\bar b \to \tau^- \tau^+$ cross section at LHC}
\label{App:2}

\begin{figure}[tbp]
\centering
  \begin{tabular}{cccc}
   \includegraphics[width=0.25\textwidth]{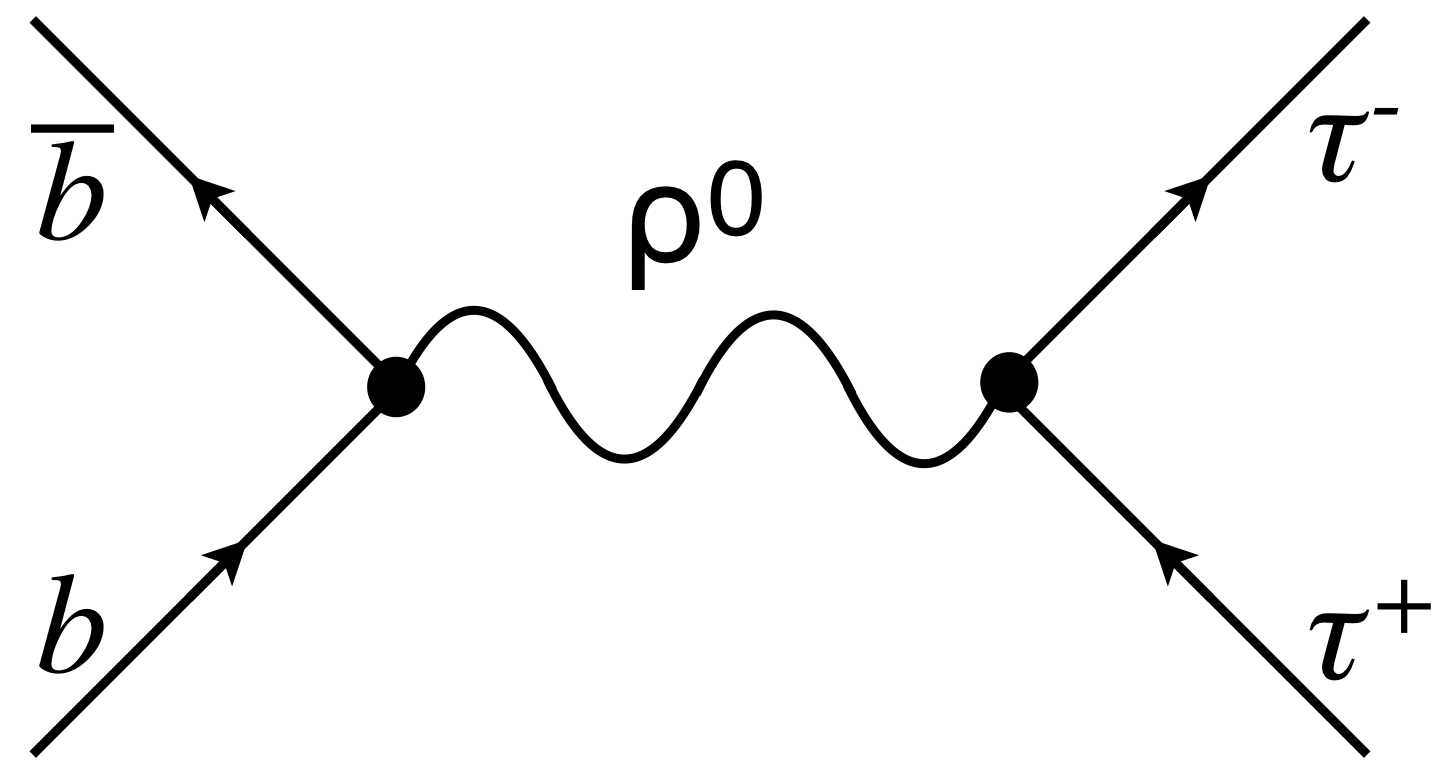} & & & \includegraphics[width=0.2\textwidth]{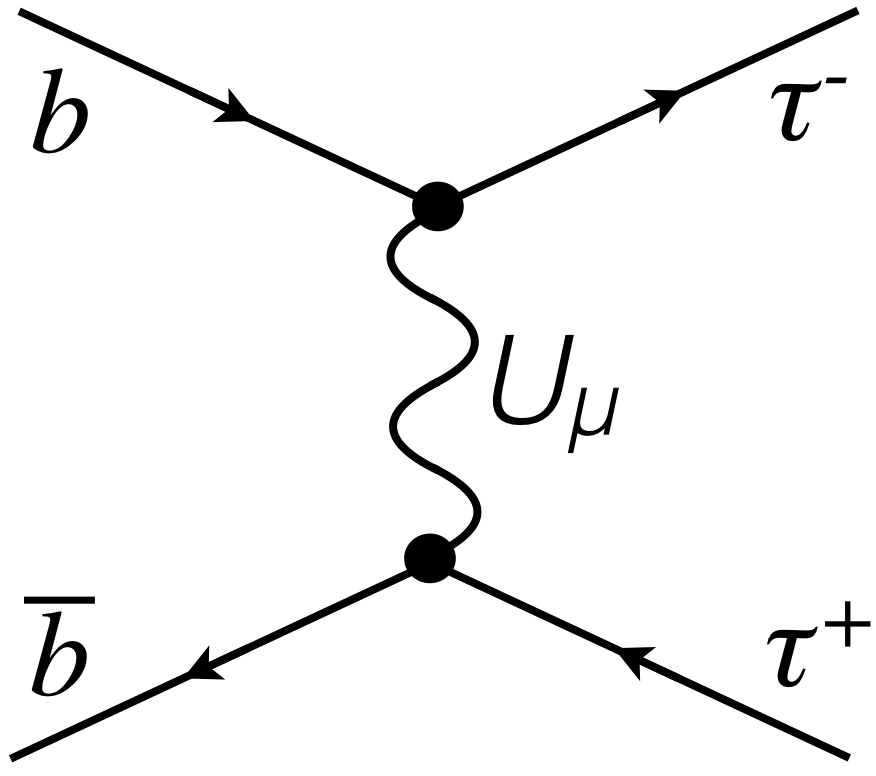}  \\ [0.5 cm]
   \end{tabular}
    \caption{ \small  Tree level diagrams for vector resonance contribution to $b ~\bar b \to \tau^- \tau^+$ production at hadron collider.
        \label{Fig:diagrams} 
        }
\end{figure}

\paragraph{$\SU(2)_L$ triplet $\rho_\mu^a$:} The relevant diagram is shown in Fig.~\ref{Fig:diagrams} (left).
The partonic cross section for $b\,\bar b \to \rho^0$ in the narrow width approximation is
\be
\sigma_N(\hat s) = \frac{4 \pi^2}{3 m_\rho} ~\Gamma(\rho^0 \to b \bar b)~\delta(\hat s - m_{\rho}^2)~,
\ee
where $\hat s$ is the partonic c.o.m. energy, while the hadronic cross section is given by
\be
\sigma(p p \to \rho^0) = 2 \mathcal{L}_{b \bar b}(\tau_\rho,\mu_F)~ \frac{4 \pi^2}{3 m_\rho s_0} ~\Gamma(\rho^0 \to b \bar b)~,
\label{eq:narrow-cross}
\ee
where $\tau_\rho = m_\rho^2/s_0$ and $\sqrt{s_0}~$ is the proton-proton c.o.m. energy. The luminosity function is defined as
\be
\mathcal{L}_{b \bar b}(\tau,\mu_F) = \int_{\tau}^1 \frac{d x}{x}~f_b (x,\mu_F) f_{\bar b} (\tau/x,\mu_F) ~,
\ee
where $f_{q}$ are the corresponding parton distribution functions.
On the other hand, the partonic cross section for $b\, \bar b \to \tau^- \tau^+$, due to the propagation of a wide $\rho^0$ resonance, is
\be
\sigma_W(\hat s) = \frac{\hat s}{2304 \pi} \left|\frac{g_q g_\ell}{\hat s-m_\rho^2 + i m_\rho \Gamma_\rho}\right|^2~,
\ee
while the hadronic one is
\be
\sigma(p p \to \tau^+ \tau^-) = 2 \int_{\tau_{\rm{min}}}^1 d \tau ~ \mathcal{L}_{b \bar b}(\tau,\mu_F)~\sigma_W(\tau s_0)~,
\label{eq:tot-cross}
\ee
where $\tau_{\rm{min}}=(m_{\tau \tau}^{\rm{min}})^2/s_0$. 
The central factorization scale is set to $\mu_F=m_\rho/2$. By inspecting more closely the narrow-width case, we find that varying the scale by a factor of two leads to a small deviation in the total cross section.  Using $68\%$ C.L. PDF sets, we also estimate the PDF uncertainty to be at the level of $\sim 20\%$.

\paragraph{Vector leptoquarks $\mathcal{U}^a_\mu$ and $U_\mu$:}  The relevant diagram is shown in Fig.~\ref{Fig:diagrams} (right). The partonic cross section for $b ~ \bar b \to \tau^- \tau^+$, due to the $t-$channel LQ exchange, is
\be
\sigma(\hat s) =\left( \frac{g_{T(S)}}{2} \right)^4 \frac{\hat s (2 + \hat s/m_{U}^2) + 2 (m_{U}^2 + \hat s) \ln (m_{U}^2/(m_{U}^2 + \hat s))}{48 \pi \hat s^2}~,
\ee
where $g_{T(S)}$ is the LQ triplet (singlet) coupling defined in Eq.~\eqref{eq:octet-qq} (Eq.~\eqref{eq:singlet-qq}).


\end{document}